\documentclass[%
 preprint, 
unsortedaddress,
amsmath,amssymb,
aps, 
physrev,
]{revtex4-2}

\usepackage{graphicx}
\usepackage{dcolumn}
\usepackage{bm}

\usepackage{algorithm}
\usepackage{algpseudocode}
\usepackage{booktabs}
\usepackage{tabularx}
\usepackage[colorlinks=true,linkcolor=blue,filecolor=magenta,
urlcolor=blue]{hyperref}
\usepackage{siunitx}
\usepackage{physics}
\usepackage{appendix}
\newcommand{\idealTetra}{109.47^{\circ}}

\begin{document}


\title{\textbf{Seamlessly joining length scales: From atomistic thermal graphs to anisotropic continuum conductivity} 
}%

\author{C. Ugwumadu}
\email{Contact author: cugwumadu@lanl.gov}
\affiliation{Quantum \& Condensed Matter (T-4) Group, Los Alamos National Laboratory, Los Alamos, 87545, NM, USA}
  
\author{D. A. Drabold}%
\affiliation{%
 Department of Physics and Astronomy, Nanoscale \& Quantum Phenomena Institute, Ohio University, 45701, OH, USA
}%

\author{R. M. Tutchton}
\affiliation{Quantum \& Condensed Matter (T-4) Group, Los Alamos National Laboratory, Los Alamos, 87545, NM, USA}%

\date{\today}

\begin{abstract}
Thermal transport in complex solids is governed by local structure, defects, and anisotropy, yet most continuum models still rely on oversimplified, homogenized conductivities. Here, we bridge atomistic and continuum descriptions by building finite element (FE) models directly from the site-projected thermal conductivity (SPTC), an atomic-level decomposition of the Green–Kubo thermal conductivity. We introduce a new toolkit, the ``Simulator Collection for Atomic-to-Continuum Scales (\texttt{SCACS})", which uses a graph neural network to predict SPTC on large atomic structures, coarse-grain these fields into anisotropic conductivity tensors, and embeds them into the heat-flow FE equation with a customized, anisotropy-aware adaptive mesh refinement scheme. Applied to silicon nanostructures, the resulting FE models act as representative volume elements, reproduce bulk conductivities, and capture interfacial and defect-driven anisotropy while maintaining thermodynamic consistency. Additionally, \texttt{SCACS} predicts experimental conductance trends and fields. This innovation demonstrates a novel and general route for transferring atomistic transport information into device-scale thermal simulations with physics-based approximations.
\end{abstract}

\maketitle




\section{Introduction}\label{sec:introduction}

Predicting heat transport in complex solids is essential across sectors where intrinsic material anisotropy and defect structures govern thermal limits and drive performance variability, including nuclear and fusion systems, semiconductor manufacturing, electronics, biotechnology, aerospace, and space applications. Structures on all scales, from atomistic to macroscopic govern heat transport. When experimental measurements are impractical or inaccessible, computational modeling offers a critical pathway for quantifying thermal behavior. Atomic-scale simulations can capture heat-transport mechanisms with high fidelity but are restricted by tractable system sizes and accessible time scales~\cite{Hu2021MultiscaleReview}. In contrast, continuum-scale approaches—particularly finite-element (FE) methods—remain the cornerstone for device-scale thermal prediction~\cite{zienkiewicz2005finite}. 

For a material with thermal conductivity tensor $\mathbf{K}(\mathbf{x})$ and a volumetric heat-generation (or removal) rate $Q$, the steady-state energy-balance requirement is:
\begin{equation}\label{eq:continuityRelation}
    \mathbf \nabla^\mathsf{T} \mathbf q(\mathbf{x}) + Q = 0  
\end{equation}
where the heat-flux vector, $\mathbf{q}$ follows Fourier’s law:
\begin{equation}\label{eq:Fourier}
    \mathbf{q}(\mathbf{x}) = -\,\mathbf{K}(\mathbf{x})\,\nabla T(\mathbf{x})
\end{equation}
with $T$ as the temperature field. In the FE solution of Equation \ref{eq:continuityRelation}, subject to prescribed boundary conditions, three quantities are central to interpreting heat transport:  
(1) the local heat flux;  
(2) the temperature gradient $\nabla T(\mathbf{x})$, which identifies thermal hotspots; and  
(3) the effective conductivity $\mathbf{K}_{\mathrm{eff}}(\mathbf{x})$, which characterizes the homogenized thermal response of the material~\cite{zienkiewicz2005finite}.

The validity of a continuum-scale FE solution depends critically on the selection of a representative volume element (RVE), which is the smallest statistically representative region of a heterogeneous material from which uniform, homogenized effective properties (e.g.,  $\mathbf{K}_{\mathrm{eff}}(\mathbf{x})$) can be extracted for use in a continuum description~\cite{Gitman2007,yvonnet2019book,Ugwumadu2024Cfoam}. For real materials, the adequacy of an RVE is assessed by examining the convergence of effective properties computed over candidate cell sizes~\cite{Gitman2007}. Two additional considerations are essential: (1) the quality of the finite-element mesh and the polynomial order $p$ of the shape functions, which together determine numerical convergence~\cite{Babuska1982}; and (2) the specification of the material properties, such as the conductivity tensor $\mathbf{K}(\mathbf{x})$ in the heat-transport problem. 

Regarding mesh quality and element order, solution accuracy improves systematically with decreasing characteristic element size $h$ and increasing polynomial order $p$ (e.g., $p=1$ for linear or $p=2$ for quadratic elements) \cite{zienkiewicz2005finite}. For second-order problems such as steady-state heat conduction, the discretization error in the energy norm typically decreases as $O(h^{p})$, while in the $L^{2}$ norm it decreases as $O(h^{p+1})$, provided the solution is sufficiently smooth. Higher-order elements therefore achieve faster convergence for a given mesh resolution, and the order of the governing PDE sets the regularity needed for the finite-element approximation to attain these optimal rates \cite{zienkiewicz2005finite}.

The nature of $\mathbf{K}(\mathbf{x})$ is more challenging to represent and is the focus of this work. In much of the FE literature, the conductivity tensor is simplified to an isotropic scalar field $k\mathbf{I}$—using either a unit value or from a bulk experimental measurement~\cite{zienkiewicz2005finite,yvonnet2019book,Yang2025,Chen2015,Meng2025}—or obtained from data-driven surrogates in which tensor components are fitted independently and inserted directly into the FE model~\cite{Xu2021SPDnn,Peng2024ML,Li2022ML,Keshav2025ML}. More general approaches adopt a diagonal tensor, $\mathbf{K}(\mathbf{x})=\mathrm{diag}(k^{x},k^{y},k^{z})$, applied at each element or integration point~\cite{Lewis2004}, or prescribe global anisotropy through directional conductivities in Cartesian coordinates~\cite{Basaula2022} or along longitudinal and transverse axes~\cite{Siddiqui2013}. These simplifications, however, cannot capture local anisotropy or sharp spatial heterogeneity, leading to errors that are often compensated by mesh refinement—adding cost and introducing mesh-dependent effective properties~\cite{Kaminski2021Homo,Schutzeichel2021Homo}.

A physically consistent representation of $\mathbf{K}(\mathbf{x})$ is required by thermodynamics. Within linear irreversible thermodynamics, the anisotropic conductivity tensor $\mathbf{K}=\{K_{ij}\}$ must satisfy two conditions: Onsager reciprocity, enforcing symmetry ($K_{ij}=K_{ji}$), and positive semidefiniteness (PSD), \(\mathbf{x}^{\top}\mathbf{K}\,\mathbf{x}\ge 0 \; \forall \; \mathbf{x}\), which guarantees non-negative entropy production~\cite{Powers}. So, when $\mathbf{K}$ is derived from homogenization or data-driven surrogates without explicitly enforcing these constraints, the result can violate thermodynamic admissibility, producing nonphysical heat-flux patterns, numerical instabilities, and mesh-dependent artifacts~\cite{Powers,Gabriel2024}. These issues motivate conductivity fields that are microscopically informed and parameterized to satisfy symmetry and PSD by construction, consistent with the \emph{direct physical approach} to finite-element analysis—rooted in atomistic rather than purely continuum considerations~\cite{zienkiewicz2005finite}—and adopted in this work.

Atomic-scale simulation has undergone a rapid transformation driven by machine-learning interatomic potentials (MLIPs). Modern MLIPs now reach density functional theory (DFT) level accuracy for energies and forces in systems far larger than those accessible to direct \textit{ab initio} or classical molecular dynamics (MD), enabling nanosecond trajectories in systems with hundreds of thousands of atoms \cite{Deringer2021,Morrow2022}. Examples include the Gaussian Approximation Potential (GAP) \cite{Bartok2010GAP}, the Hierarchically Interacting Particle Neural Network (HIP-NN) \cite{Lubbers2018hipNN}, and the Multi-scale Atomic Cluster Expansion (MACE) \cite{Batatia2025}, among others. In terms of configuration space and timescale, the atomic scale is no longer the primary bottleneck. 

By contrast, conventional atomistic thermal-transport methods often deliver effective, cell-averages thermal thermal conductivity values. For example, the Green--Kubo formula (based on equilibrium MD) returns a single bulk conductivity value obtained from heat-current autocorrelations \cite{Kubo_TC,Green_TC,Schelling2002Comparison,Kang2017GKF,JJ2021}. Direct non-equilibrium MD imposes a temperature gradient or heat baths to extract an effective conductivity along a prescribed direction \cite{McQuarrie1997,MuellerPlathe1997NEMD,Zhou2015NEMD,Maginn2018EMD}, and the Boltzmann transport equation or modal analyses provide directional and spectral conductivities for crystalline or weakly disordered solids \cite{Broido2007BOLTZ, Koh2007,Yang2013MFP,Larkin2014MFP,Lindsay2016BOLTZ}. In all these cases, the output is essentially a homogenized conductivity (or a small set of direction-dependent values) for the simulated cell, with limited information provided about how different atomic environments, interfaces, or defects contribute locally to heat transport.

The Site-Projected Thermal Conductivity (SPTC; $\zeta$) addresses this gap by recasting the Allen–Feldman implementation of the Green–Kubo equation into atomic-site contributions \cite{AF1,AF3,Ugwumadu2025}. These site-projected conductivities retain directional anisotropy, phase boundaries, and defect-mediated pathways, and can be coarse-grained over arbitrary regions to recover both conventional effective conductivities and spatial maps of transport channels. SPTC has been successfully applied to silicon and silicon-germanium systems \cite{Ugwumadu2025,Gautam2025} , graphene \cite{Gautam2025}, tungsten \cite{UgwumaduPSSB2025}, and the phase-change germanium–antimony–telluride (Ge$_2$Sb$_2$Te$_5$) chalcogenide glass \cite{Nepal2025GST}. However, current implementations require force-constant and dynamical matrices and involve eigenmode operations whose cost grows steeply with system size, so practical applications have so far been limited to models of order $10^3$–$10^4$ atoms \cite{Ugwumadu2025}.

This work handles the limitations in how the conductivity tensor is represented in finite element methods (FEM) for heat flow by introducing an atomic-level, anisotropic conductivity field learned from SPTC and upscaled into an FE model. We introduce the \emph{Simulator Collection for Atomic-to-Continuum Scales} (\texttt{SCACS}), an end-to-end framework for atom-informed FEM modeling of heat transport. 

\texttt{SCACS} comprises three main components. First, the \texttt{sptc-ai} module is a graph neural network (GNN) predictor that learns structure–transport relations from smaller SPTC-labeled systems and generalizes them to the massive ($\gtrsim 10^5$ atoms), near device-scale structures now accessible with MLIPs, accelerating SPTC analysis by orders of magnitude. Second, the \texttt{sptc2fem} module upscales the true or predicted SPTC into a smooth conductivity tensor field via Gaussian smoothing and structured voxelization, and maps this field onto FE meshes as element- and node-based conductivity tensors $\mathbf{K}(\mathbf{x})$, yielding RVEs equipped for Galerkin discretization. By construction, this implementation ensures symmetry and PSD consistent with the SPTC formula.

Third is the \texttt{python}-based FEM solver implemented using the Unified Form Language (UFL) and the \texttt{DOLFINx} framework within the FEniCS project~\cite{UFL,DOLFINx}. The steady heat-flow problem is discretized in this framework and solved using the Portable, Extensible Toolkit for Scientific Computation (PETSC) library \cite{PETSc}. This then yield the temperature fields, heat-flux maps, and anisotropic flux-based effective conductivity $K_\mathrm{eff}^\alpha$ along the thermal loading direction, $\alpha$ and under prescribed boundary conditions. Within this workflow we also introduce an Equal-Mass Controller (EMC), a structured-grid adaptive mesh refinement (AMR) strategy driven by a residual-based flux-imbalance indicator that redistributes grid planes toward high-error regions, avoids hanging nodes~\cite{zienkiewicz2005finite,yvonnet2019book,Bangerth2017}, and maintains compatibility with the voxelized, lattice-aligned SPTC values.

As a case study, we deploy \texttt{SCACS} on silicon (Si), a cornerstone material in nanoscale and device engineering, ubiquitous in thermoelectric, microelectronic and microelectromechanical systems (MEMS), where thermal design is critical \cite{Uma2001,Abel2007,hochbaum2008nanowires,Boukai2008nanowires,Takeuchi2020PolySi,DavisNanopillars2014}, and whose SPTC properties have been extensively studied \cite{Ugwumadu2025,Gautam2025}. We train an equivariant GNN on carefully curated Si supercells containing $\mathcal{O}(10^3)$ atoms each, spanning crystalline, polycrystalline, paracrystalline, and amorphous structures. The model predicts per-atom SPTC Cartesian components $\hat{\zeta}^x,\,\hat{\zeta}^y,\,\hat{\zeta}^z$, as well as an isotropic value $\hat{\zeta}^\mathrm{d}$, that derive from local bonding topology and chemistry. These fields are then coarse-grained into spatially varying, anisotropic $\mathbf{K}(\mathbf{x})$. The resulting FE solutions capture interfacial behavior and defect-mediated anisotropy in a thermodynamically consistent way, transferring atom-scale transport information into continuum-scale temperature and heat-flux fields while preserving both material fidelity and realistic device geometry.

\section{The Site-Projected Thermal Conductivity}

Within linear-response theory, transport coefficients follow from the fluctuation–dissipation theorem (Kubo formula), which relates them to time integrals of equilibrium current autocorrelations~\cite{Kubo_TC}. In classical MD, the heat-flux autocorrelation function (ACF) can be evaluated directly via the Hardy current \cite{Hardy} and integrated to obtain the thermal conductivity \cite{Uguwmadu2024NPC}. These simulations naturally include anharmonic effects but become unreliable at low temperatures where quantum statistics are dominate.

Allen and Feldman (AF) addressed this limitation by reformulating the problem within the harmonic approximation for a quantized lattice, deriving a Green–Kubo expression \cite{Kubo_TC,Green_TC} for the thermal conductivity that depends on the force-constant matrix, dynamical eigen-pairs, atomic masses, and equilibrium coordinates~\cite{AF1,AF3}. At the Brillouin-zone center ($\Gamma$; $\mathbf{k}=0$), where normal-mode eigenvectors are real, the AF formulation can be reordered to express the directional thermal conductivity ($K^\alpha_\zeta$) as a double sum of a real-symmetric thermal matrix $\Xi_{ij}^\alpha$ over atomic sites ($i,j$) along a given Cartesian component $\alpha=x,\;y,\text{or}\;z$, as~\cite{Ugwumadu2025}:

\begin{equation}\label{eq:AF_kappa}
    K^\alpha_\zeta \;=\;  \sum_{i}  \;  \zeta^\alpha_i \;\equiv\; \sum_{i,j} \, \Xi^\alpha_{ij} 
\end{equation}
where:
\begin{align}
\label{eq:AF_Xi}
\Xi^\alpha_{ij} =\;&
\frac{\pi \hbar^2}{48\,k_B T^2 V}\;
\sum_{p\neq q}
\delta(\omega_p-\omega_q)\,
\frac{(\omega_p+\omega_q)^2}{\omega_p\omega_q}\,
\frac{e^{\hbar\omega_p/k_BT}}{\big(e^{\hbar\omega_p/k_BT}-1\big)^2}\;\;
\sum_{\gamma,\beta}\frac{\phi^{\alpha\beta}_{ij}(0,\gamma)}{\sqrt{m_i m_{j}}}   \,e^{\alpha p}_{i,0}\,e^{\beta q}_{j,0}\\[-2pt]
&
\sum_{\gamma'}\sum_{a,b}\sum_{\alpha',\beta'}
\frac{\phi^{\alpha'\beta'}_{ab}(0,\gamma')}{\sqrt{m_a m_b}}\,
e^{\alpha' p}_{a,0}\,e^{\beta' q}_{b,0}\,\,
\sum_{\eta}\big(R^\eta_{\gamma}+R^\eta_{ab}\big)\,\big(R^\eta_{\gamma'}+R^\eta_{ij}\big)
\notag
\end{align}
and $V$ is the cell volume, $m_i$ the atomic mass, $T$ the temperature, $\phi^{\alpha\beta}_{ij}$ the force-constant tensor, $e^{\alpha p}_{i,0}$ the $\alpha$ polarization of mode $p$ at site $i$, and $\omega_p$ the mode frequency. $R^\eta_{\gamma}$ and $R^\eta_{ij}$ denote the $\eta$ component of the $\gamma$-th cell vector and the displacement between sites $i$ and $j$, respectively.  

The symbol $\zeta$ denotes the site-projected thermal conductivity (SPTC). Equation~\ref{eq:AF_kappa} gives the anisotropic total conductivity, where the projection onto atom $i$, written as $\zeta_i^\alpha$, is the directional SPTC along Cartesian direction $\alpha$. The isotropic (directionally averaged) quantities, $K_\zeta^{\mathrm{d}}$ and $\zeta_i^{\mathrm{d}}$, are then obtained by averaging over the three Cartesian components:

\begin{equation}\label{eq:SPTC_zeta}
    K^\mathrm{d}_\zeta \;=\;  \sum_{i}  \;  \zeta^\mathrm{d}_i \;\equiv\;  \sum_{i,\alpha} \frac{\zeta^\alpha_i }{3}
\end{equation}

Built entirely from lattice-dynamical quantities (force constants, phonon frequencies and eigenvectors, and lattice translations), $\Xi$ satisfies key structural properties: it is real symmetric (enforcing Onsager reciprocity), admits an eigen-decomposition: \(\Xi\,\psi_\mu = \lambda_\mu\,\psi_\mu,
\) and \( \sum_\mu \lambda_\mu = \mathrm{tr}\,\Xi = 0 \). Additionally, $\Xi$ decays rapidly with interatomic separation $|\mathbf{r}-\mathbf{r}'|$ due to the locality of force constants~\cite{Ugwumadu2025}. These features enable controlled spatial averaging of the SPTC fields into element-wise anisotropic conductivity tensors for FE implementation while automatically preserving symmetry and enforcing PSD.


\section{The \texttt{sptc-ai} Framework}
The training and validation datasets were selected from a pool of 653 Si structures drawn from the \emph{P}, \emph{Q}, and \emph{R} groups (Section~\ref{subsubsec:models_training}). Selection was guided by an analysis of the atomic-level descriptors used to construct the node and edge features. The directed edge feature set includes a 32-dimensional Gaussian-type radial basis expansion of the interatomic distance and a three-component unit displacement vector evaluated under the minimum-image convention (Section~\ref{subsubsec:features}).

For node (atom) features, we defined six continuous structural descriptors: coordination number (CN), the root-mean-square deviation from the ideal tetrahedral angle ($\angle\mathrm{rms}$), the mean ($B_{\mu}$) and standard deviation ($B_{\sigma}$) of bond-lengths, and the fractions of short ($B_{\mathrm{S}}$) and long ($B_{\mathrm{L}}$) bonds. A 64-dimensional principal-component representation of the Smooth Overlap of Atomic Positions (SOAP) descriptor~\cite{SOAP} was also included (Section~\ref{subsubsec:features}).

Local atomic environments were further classified using the ``Identify Diamond'' modifier in OVITO~\cite{OVITO}. This procedure assigns atoms to amorphous or crystalline categories, with crystalline atoms distinguished by their first- and second-nearest-neighbor shells (1-NN and 2-NN) as either cubic diamond (3C) or hexagonal diamond (2H). The labeling scheme is: ID~0—amorphous; IDs~1 and~4—perfect 3C and 2H; IDs~2–3 (3C) and~5–6 (2H)—first- and second-nearest neighbors, respectively. Additional details on the models and feature construction, including global descriptors, are provided in Section~\ref{subsec:sptc_ai} and in Section~\textcolor{blue}{S1} of the supplemental Material \cite{SM}.

\begin{figure}[!b]
    \centering
    \includegraphics[width=\linewidth]{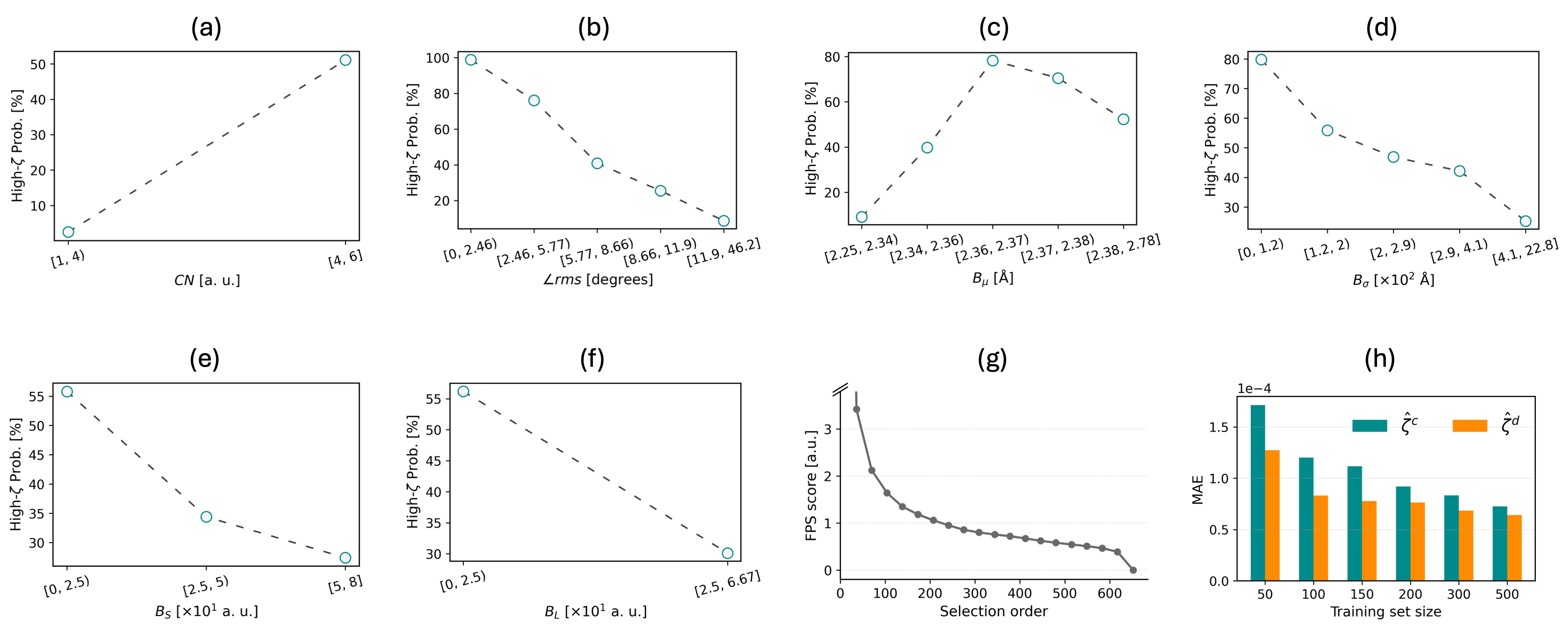}
    \caption{\textbf{Architecture and Training of \texttt{sptc-ai}.} Empirical probability of high-SPTC outcome from binned distribution of the continuous structural descriptors, including (a) the coordination number (CN), (b) the root-mean-square deviation from the ideal tetrahedral angle ($\angle\mathrm{rms}$), (c) the mean (B$_\mu$) and (d) standard deviation (B$_\sigma$) of bond-lengths, and the (e) fractions of short (B$_\mathrm{S}$) and (f) long (B$_\mathrm{L}$) bonds. (g) Farthest point sampling score for all structures. (h)  Mean absolute error (MAE) for the averaged directional SPTC ($\hat{\zeta}^\mathrm{c}$) and the directly predicted isotropic SPTC ($\hat{\zeta}^\mathrm{d}$).}
    \label{fig:Fig_sptcai_architecture}
\end{figure}

The SPTC values for the \emph{P}, \emph{Q}, and \emph{R} structures exhibit right-skewed distributions with long high-SPTC tails (Table~\ref{tab:SPTCStats}). To identify factors contributing to high SPTC and to mitigate potential majority-class bias in training the GNN model, we analyzed the marginal distributions of six continuous structural descriptors together with the corresponding probability of high-SPTC atoms (Figure~\ref{fig:Fig_sptcai_architecture}a--f). Following the approach of Reference~\cite{Ugwumadu2025}, we define a binary label by selecting atoms in the upper and lower tails of $\zeta_i^{\mathbf d}$:
\begin{equation}
Y(i)=
\begin{cases}
1,& \zeta^\mathbf{d}_i \ge \mathcal{Q}_{0.8}(\zeta)\quad\text{(high SPTC)}\\[2pt]
0,& \zeta^\mathbf{d}_i \le \mathcal{Q}_{0.2}(\zeta)\quad\text{(low SPTC)}
\end{cases}
\end{equation}
where $\mathcal{Q}_p(\zeta)$ denotes the $p$-quantile of the empirical distribution of $\{\zeta_i^{\mathbf d}\}$ (e.g., $\mathcal{Q}_{0.8}$ is the 80th percentile). Atoms with $\mathcal{Q}_{0.2}(\zeta) < \zeta_i^{\mathbf d} < \mathcal{Q}_{0.8}(\zeta)$ (the middle 60\% of SPTC values) are excluded from this classification. The association between each descriptor and the high/low SPTC label is quantified using a relative-risk metric~\cite{ranganathan2015}:
\begin{equation}\label{eq:RR}
\mathrm{RR} = \frac{a/(a+b)}{c/(c+d)}
\end{equation}
where $(a,b)$ denote the high/low counts in the top descriptor bin and $(c,d)$ denote the corresponding counts in the bottom bin. To avoid zero-cell instabilities, we apply the Haldane--Anscombe correction \cite{Haldane1956,Anscombe1956} by adding $0.5$ whenever $\min(a,b,c,d)=0$.

\begin{table}[!t]
    \caption{SPTC statistics for the Si dataset. Values are scaled by $\times 10^{-3}$ $\mathrm{W\,m^{-1}\,K^{-1}}$.}
    \centering
    \scriptsize
    \begin{tabular}{l*{7}{S[table-format=3.4]}}
    \toprule
    {Prefix} & {min} & {p05} & {p25} & {median} & {p75} & {p95} & {max} \\
    \midrule
    P   & 0.1002 & 0.3734 & 0.5174 & 0.6311 & 0.7501 & 0.9669 & 2.624 \\
    Q   & 2.566 & 9.022 & 10.80 & 12.01 & 13.24 & 16.24 & 17.09 \\
    R   & 0.06287 & 0.4922 & 0.6226 & 0.7139 & 0.8119 & 1.115 & 3.341 \\
    All & 0.1031 & 0.5466 & 0.7116 & 0.8271 & 0.9719 & 4.385 & 17.0900 \\
    \bottomrule
    \end{tabular}
    \label{tab:SPTCStats}
\end{table}

The coordination number strongly influences SPTC (Figure~\ref{fig:Fig_sptcai_architecture}a). Under-coordinated environments (CN$<4$) are associated with low SPTC, whereas well-coordinated Si environments (CN~$\ge 4$) have $\mathrm{RR}\approx 3.86$, corresponding to an  $\approx 3.9$-fold higher likelihood of high SPTC. Angular order ($\angle\mathrm{rms}$) shows an equally pronounced effect (Figure~\ref{fig:Fig_sptcai_architecture}b): the high-$\zeta$ fraction decreases monotonically from $\sim 0.94$ in the most ordered bin to $\sim 0.19$ in the  most disordered bin, indicating a strong protective role of low angular distortion.

Bond-length descriptors show subtler effects. The mean bond-length $B_{\mu}$ lies in the range 2.36--2.39~\AA, with an overall $\mathrm{RR}\approx 1.27$ (Figure~\ref{fig:Fig_sptcai_architecture}c). From the per-group distributions in Figure~\textcolor{blue}{S1} \cite{SM}, the \emph{P} and \emph{R} structures exhibit a stronger dependence on $B_{\mu}$ (RR~$\approx 2.2$), consistent with the well-known sensitivity of SPTC to topological disorder, defects, and grain boundaries~\cite{Ugwumadu2025,Gautam2025,UgwumaduPSSB2025,Nepal2025GST}. In contrast, the \emph{Q} models (predominantly crystalline) show an almost flat response centered near $2.35$~\AA, the equilibrium Si bond-length.
Bond-length dispersion ($B_{\sigma}$) is consistently unfavorable (Figure~\ref{fig:Fig_sptcai_architecture}d), as are the short- and long-bond tail fractions (Figures~\ref{fig:Fig_sptcai_architecture}e--f), both of which correspond to markedly reduced probabilities of high SPTC.

The next step was to ensure structural diversity in the training set. For this, we constructed quantitative embeddings ($s_i$) for each structure. A feature vector $\mathbf{x}_i \in \mathbb{R}^{24}$ was constructed by aggregating statistics of the aforementioned descriptors. This vector include the SPTC quantiles and standard deviation, analogous mean–standard-deviation pairs for the descriptors, and structure class. Each feature, $i$ was standardized to zero mean and unit variance, yielding a normalized embedding $\mathbf{z}_i = (\mathbf{x}_i - \boldsymbol{\mu}) / \boldsymbol{\sigma}$.

A representative subset of structures for training the GNN were then selected following farthest-point sampling (FPS)~\cite{Gonzalez1985FPS} in this standardized feature space. At each iteration, FPS selects the structure whose embedding lies farthest from those already chosen—that is, the structure whose nearest neighbor among the selected set is farther away than that of any other unselected structure as:
\begin{equation}
\min_{i \in S_t} \lVert \mathbf{z}_{s_{t+1}} - \mathbf{z}_i \rVert_2 
\ge
\min_{i \in S_t} \lVert \mathbf{z}_{j'} - \mathbf{z}_i \rVert_2
\end{equation}
where $S_t$ is the set of previously selected samples. The algorithm is initialized with the structure having the largest Euclidean norm, corresponding to the most extreme configuration in standardized space. The resulting distribution of structural classes for the top 100 selected training structures is shown in Figure~\textcolor{blue}{S2}a with colors indicating the structure class IDs~0–6 \cite{SM}, and corresponding FPS ordering in Figure~\ref{fig:Fig_sptcai_architecture}g. The FPS for the 2nd, 100th and 653rd Si structure is $\approx$ 28.2, 1.7 and 0.001, respectively, while the first structure (the seed) has FPS $=0$ by construction (self-similar). 

Based on the FPS analysis, we performed stability tests for the GNN underlying \texttt{sptc-ai}, a Graph Isomorphism Network (GIN) implemented using the \texttt{GINEConv} module in PyTorch Geometric~\cite{GINEConvpaper,GINEConvpytorch}. The architecture is described in Section~\ref{subsubsec:sptc_ai_gin}, and a schematic is provided in Figure~\textcolor{blue}{S2}b. Figure~\ref{fig:Fig_sptcai_architecture}h shows the mean absolute error (MAE) in predicting the averaged directional SPTC ($\hat{\zeta}^\mathrm{c}$) and the directly predicted isotropic SPTC ($\hat{\zeta}^\mathrm{d}$) as a function of training-set size along the FPS ordering. The first 500 structures define the candidate training pool; the next 120 structures are held out as an external test set, and the remaining 33 low-FPS structures are excluded. Overall, the GNN exhibits strong predictive performance for all training subsets, with the MAE saturating for training sets larger than about 200 structures. In this work we adopt the 300-structure training subset, which yields a well-converged training loss (Figure~\textcolor{blue}{S2}c) and a tight correlation between computed SPTC values and model predictions on the held-out test set (Figure~\textcolor{blue}{S2}d).

\begin{figure}[!t]
    \centering
    \includegraphics[width=\linewidth]{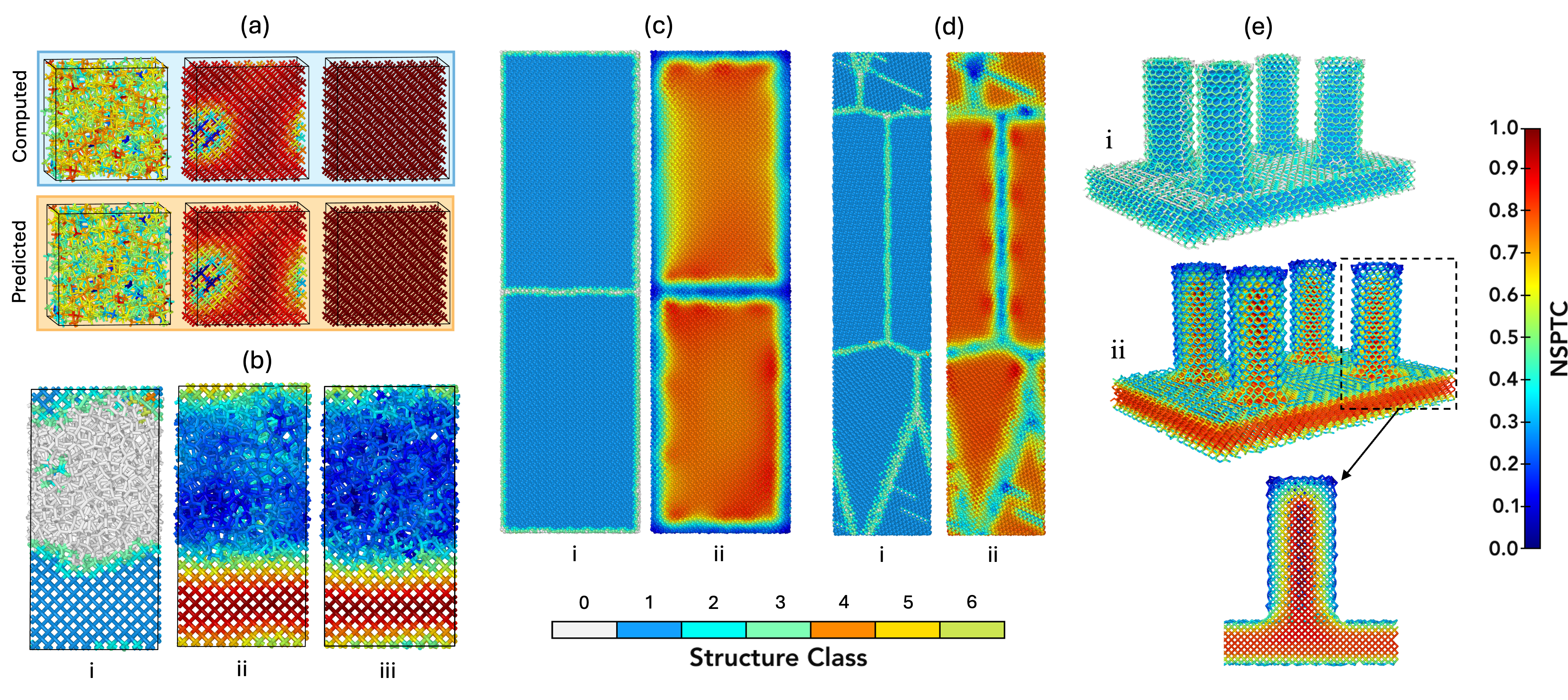}
    \caption{\textbf{Performance of \texttt{sptc-ai}.} 
    The ``NSPTC'' colorbar indicates the normalized SPTC, ranging from 0 (blue) to 1 (red), while the ``Structure class'' colorbar encodes the OVITO-based local environment: amorphous (0; gray), cubic diamond (1; blue) and its first (2; cyan) and second (3; green) neighbors, and hexagonal diamond (4; orange) with its first (5; yellow) and second (6; lime) neighbors.  
    (a) Model performance on held-out test structures: computed (top) and predicted (bottom) SPTC distributions 
    for (left$\rightarrow$right) amorphous, defective, and purely crystalline Si. 
    (b) Generalization of \texttt{sptc-ai} to an unseen amorphous–crystalline Si interface containing 2022 atoms, showing (i) structure class, (ii) predicted SPTC, and (iii) computed SPTC. 
    (c--e) Structure class and predicted SPTC for (c) a 223,930-atom Si twin-grain-boundary nanowire, (d) a 134,923-atom polycrystalline Si nanowire, and (e) a 26,191-atom Si nanopillar. Panels (c) and (d) show mid-slices; the black rectangle in (e) highlights the zoomed mid-slice region of the nanopillar.}
    \label{fig:Fig_sptcai_models}
\end{figure}

Figure~\ref{fig:Fig_sptcai_models}a compares model predictions for three representative Si systems (left to right): an amorphous structure, a defective crystal, and a fully crystalline structure. The agreement with the computed SPTC is highest for the crystalline case (reflecting its uniform lattice ordering) and lowest for the amorphous one. The ``NSPTC'' colormap denotes the per-atom normalized SPTC values. For additional validation, we applied \texttt{sptc-ai} to an unseen 2022-atom amorphous–crystalline (AC) Si interface whose SPTC characteristics have been studied previously~\cite{Ugwumadu2025}. The structural class analysis for the AC model is shown in Figure~\ref{fig:Fig_sptcai_models}b[i]. The predicted SPTC field
(Figure~\ref{fig:Fig_sptcai_models}b[ii]) closely follows the computed SPTC (Figure~\ref{fig:Fig_sptcai_models}b[iii]) across the interface and reproduces the overall variation within the amorphous region.

For the finite-element implementation, in addition to the AC structure, we applied the validated GNN model to three larger, structurally realistic Si nanostructures that are relevant to microelectronic, photonic, thermoelectric, and related technologies~\cite{hochbaum2008nanowires, Boukai2008nanowires,Takeuchi2020PolySi,DavisNanopillars2014}, and for which direct SPTC analysis is infeasible: a 223{,}930-atom twin-grain-boundary structure, a 134{,}923-atom polycrystalline
nanowire, and a 26{,}191-atom nanopillar (Figure~\ref{fig:Fig_sptcai_models}c--e). For each case, panel~[i] shows the OVITO-based structural classification, and panel~[ii] shows the predicted SPTC field. The atomic structure files with SPTC predictions, in \texttt{.xyz} format, are provided as supplemental data \cite{Zenodo}.

\section{The \texttt{sptc2fem} Framework}\label{subsec:results_sptc2fem}
The key ingredient of the finite element form of the constitutive relation for the heat-flow problem is the conductivity tensor $\mathbf{K}(\mathbf{x})$ (see Section~\ref{subsubsec:heatEquation}). The \texttt{sptc2fem} module in SCACS maps the atomic-scale SPTC, interpreted as $\mathbf{K}(\mathbf{x})$, onto a regular voxel grid used in the finite element heat-flow equations, thus creating suitable RVEs from which the effective conductivity, $\mathbf{K}_\mathrm{eff}$ is obtained (see implementation in Section~\ref{subsubsec:voxelization} and \ref{subsubsec:Keff}). The FE mesh structure (in \texttt{.xdmf} and associated \texttt{.h5} format) for all models discussed herein are provided as supplemental data \cite{Zenodo}.

\begin{figure}[!t]
    \centering
    \includegraphics[width=\linewidth]{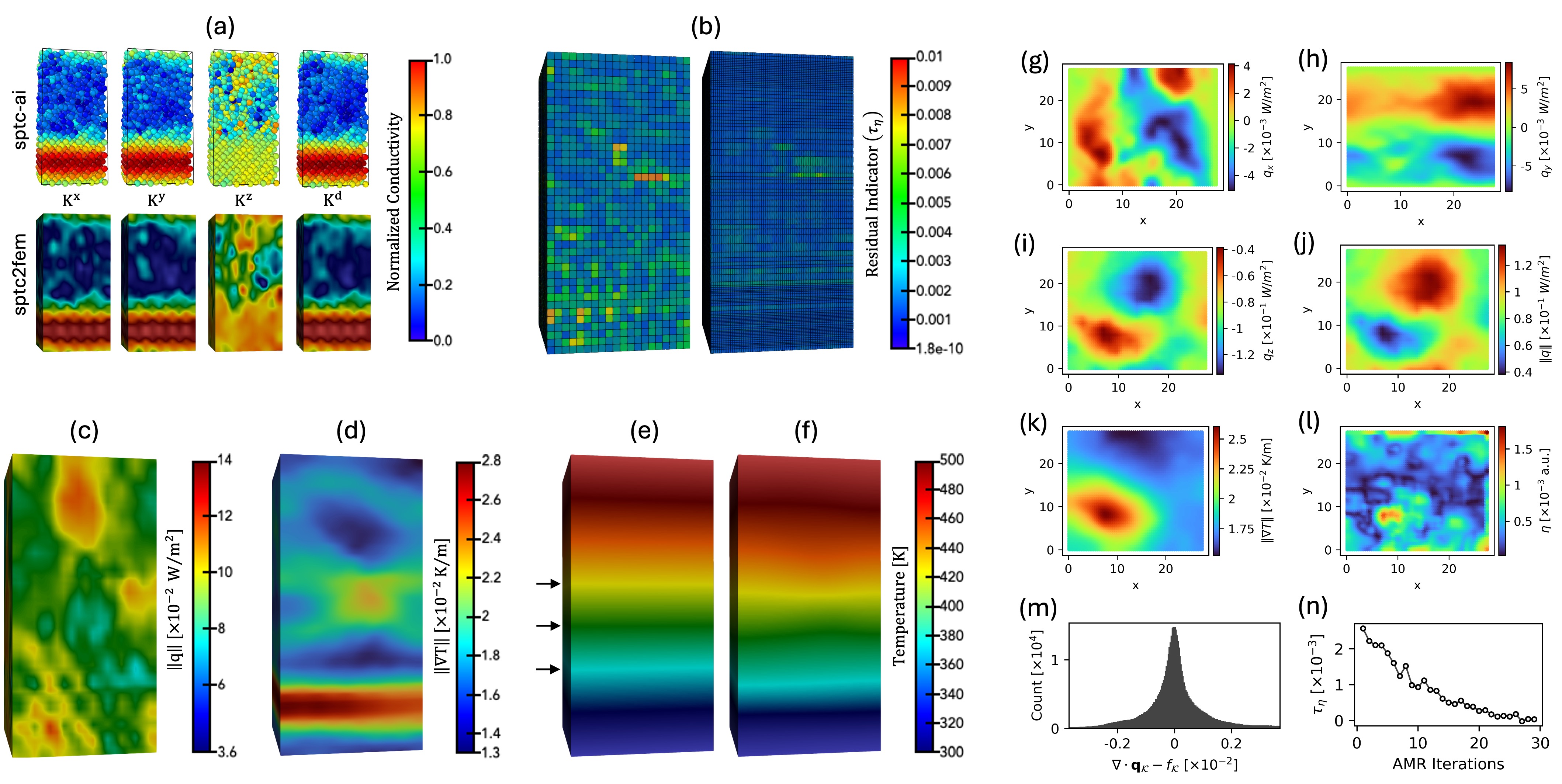}
    \caption{\textbf{Analysis of \texttt{sptc2fem} results for the amorphous–crystalline Si structure.}
(a) Anisotropic and isotropic SPTC (top) and coarse-grained conductivity (bottom); colormap shows normalized values. 
(b) Adaptive mesh refinement using the equal mass controller with D{\"o}rfler marking, showing coarse (left) and refined (right) meshes. 
(c, d) Dirichlet solve for thermal loading along $z$: magnitude of the temperature gradient and flux. 
(e) steady-state temperature for $\mathbf{K}(\mathbf{x})= \mathbf{1}$ and (F) for \texttt{sptc-ai}; arrows mark regions with differing temperature transitions. 
(f--l) Slices at $z = 30$ \AA{} of directional fluxes, flux magnitude, temperature-gradient magnitude, and flux residuals $\eta$. 
(m) Histogram of continuity residual and (n) the AMR convergence.
    }
    \label{fig:Fig_sptc2fem_AC}
\end{figure}

The anisotropic and isotropic SPTC and the corresponding conductivity fields for the AC structure are shown in the top and bottom panels of Figure~\ref{fig:Fig_sptc2fem_AC}a, respectively. The $\mathbf{K}(\mathbf{x})$ mirrors the $\zeta$ mapping even at the amorphous-crystalline interface. Achieving this required refining an initial coarse mesh that under-resolves these features using the heat-flux-aware adaptive mesh refinement (AMR) procedure that implements a residual-based ($\eta$) error indicator and employs a D{\"o}rfler-type bulk marking strategy~\cite{Dorfler1996} to identify high-error regions, concentrate refinement in regions of large $\eta$, and regenerate the mesh (global $h$-refinement) using the novel Equal-mass controller (EMC) (see discussion in Section~\ref{subsubsec:amr-sptc}).  The starting (coarse) and final (refined) mesh for the AC structure is shown in left and right panel of Figure \ref{fig:Fig_sptc2fem_AC}b, respectively. The black lines indicates the grids and the colormap indicate the  $\eta$ values for each cell, $\mathcal{K}$.

Unless otherwise stated, we solve the standard steady-state heat conduction problem with Dirichlet boundary conditions, prescribing the temperature on the two faces normal to the thermal drive direction ($x$, $y$, or $z$). The magnitude of the heat flux, $\|\mathbf{q}\|$, shows variation within the RVE (Figure~\ref{fig:Fig_sptc2fem_AC}c), while the magnitude of the resulting temperature gradient $\|\nabla T\|$ for the AC structure along the elongated $z$-axis also shows a noticeable change across the amorphous–crystalline interface (Figure~\ref{fig:Fig_sptc2fem_AC}d). Together, these fields directly reflect the anisotropy encoded in $\mathbf{K}(\mathbf{x})$ and the resulting effective conductivity obtained from homogenization. In conventional FE analysis, resolving such localized gradients typically requires introducing artificially complex geometries; in contrast, \texttt{SCACS} achieves this level of detail on simple domains (such as the cuboid considered here), reducing manual geometric tuning and the associated user bias.

The impact of the SPTC-based conductivity field is also evident in the temperature profiles implied by the Dirichlet boundary conditions. In Figure~\ref{fig:Fig_sptc2fem_AC}e, the reference solution with $\mathbf{K}(\mathbf{x}) = \mathbf{1}$ shows an almost linear transition between the end facets, whereas the \texttt{SCACS}-based solution in Figure~\ref{fig:Fig_sptc2fem_AC}f displays pronounced nonlinear curvature in the colormap (black arrows). This curvature signals directional anisotropy and highlights specific regions of the RVE that either enhance or impede thermal transport.

We show a slice in the $x$–$y$ plane of the directional flux components $q^\alpha$, the flux magnitude $\|\mathbf{q}\|$, and the temperature–gradient magnitude $\|\nabla T\|$ at $z = 30$~\AA{} for the same $z$-directed thermal solve (Figure~\ref{fig:Fig_sptc2fem_AC}g--l). The in-plane flux components $q^x$ and $q^y$ are nonzero on this slice due to the anisotropic and spatially varying conductivity field $\mathbf{K}(\mathbf{x})$ (Figures~\ref{fig:Fig_sptc2fem_AC}g,h), but they are more than an order of magnitude smaller than $q^z$ (Figure~\ref{fig:Fig_sptc2fem_AC}i), along which
the heat-flow problem is driven. Consequently, $q^z$ dominates the flux magnitude $\|\mathbf{q}\|$ (Figure~\ref{fig:Fig_sptc2fem_AC}j). Additionally, $\|\nabla T\|$ highlights regions of sharp temperature variation and potential thermal bottlenecks, particularly where the local heat flux is small, which is a useful diagnostic for many thermal devices (Figure~\ref{fig:Fig_sptc2fem_AC}k).

The dominant contributors to large local values of the element-wise residual indicator $\eta$ are interfaces, voids, defects, strongly anisotropic regions, and, in some cases, boundary (edge) effects. In the slice shown here, the largest $\eta$ values are concentrated near the exterior boundary of the structure (Figure~\ref{fig:Fig_sptc2fem_AC}l), consistent with edge effects commonly observed in both atomistic and continuum simulations. The distribution of the residual of the continuity equation for energy balance (Equation~\ref{eq:continuityRelation}), shown in Figure~\ref{fig:Fig_sptc2fem_AC}m, is sharply peaked near zero with only weak tails, indicating that the steady-state energy balance is satisfied up to small numerical errors.

Since $\eta$ decrease under mesh refinement but do not vanish exactly, we adopt two stopping criteria for adaptive refinement: the flux-based effective conductivity and a global residual indicator (built from $\eta$) must both stabilize between consecutive refinement levels within tolerances $\tau_\mathrm{K}$ and $\tau_\eta$, respectively (Equation~\ref{eq:AMRcirteria}).  The convergence of the global residual indicator toward the tolerance $\tau_\eta$ for the AC structure is shown in Figure~\ref{fig:Fig_sptc2fem_AC}n. The corresponding initial and final meshes in this refinement sequence are illustrated in Figure~\ref{fig:Fig_sptc2fem_AC}b.

\begin{figure}[!tbhp]
    \centering
    \includegraphics[width=\linewidth]{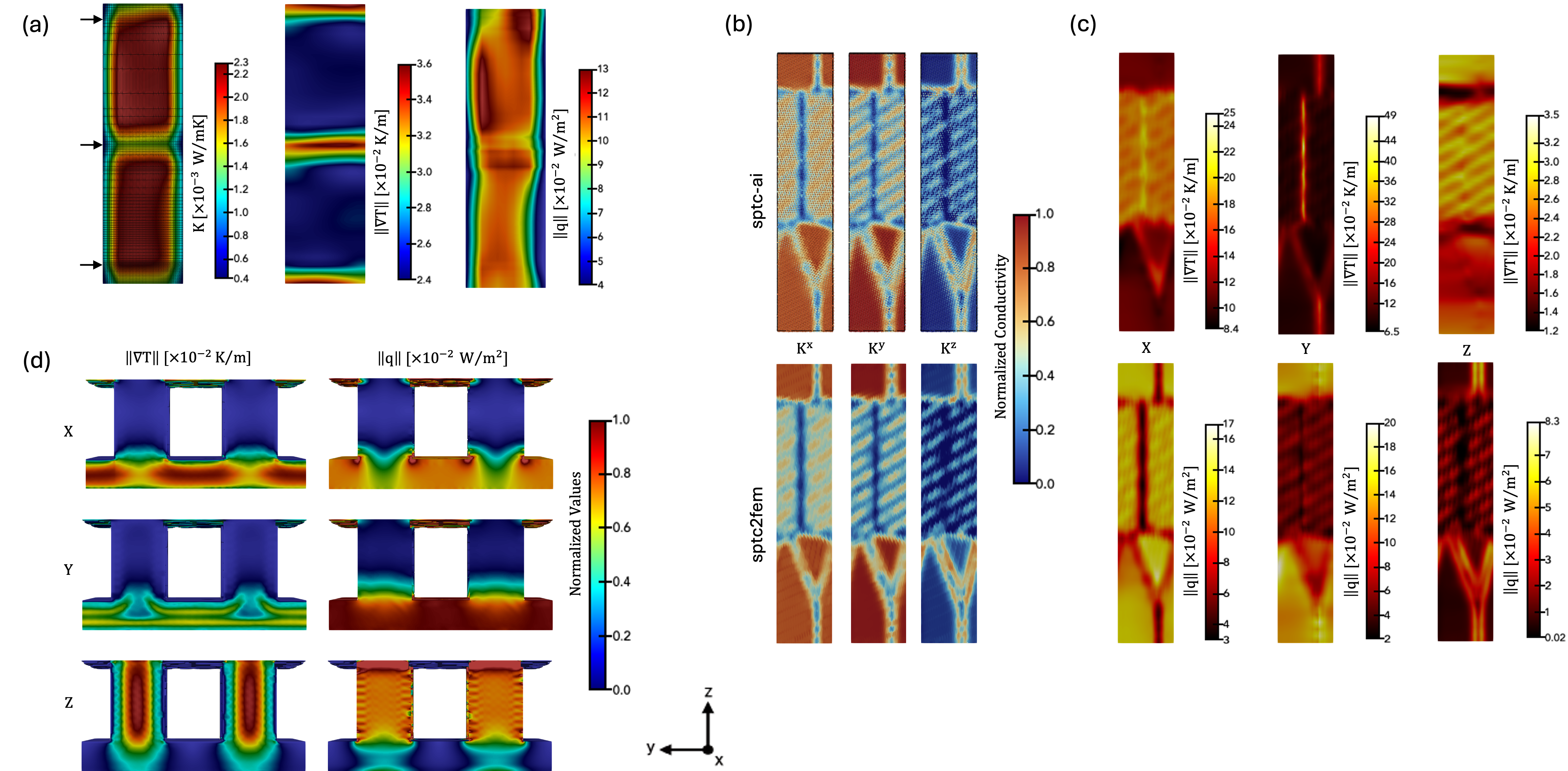}
    \caption{\textbf{Analysis of the \texttt{sptc2fem} results for the other structures.} 
(a) Twin grain boundary: isotropic conductivity (arrows mark EMC-refined regions), temperature gradient, and flux (left to right). 
(b) Polycrystalline Si: anisotropic SPTC (top) and coarse-grained conductivity (bottom); colormap shows normalized values, and
(c) the magnitude of temperature gradient (top) and flux (bottom) for thermal loading along $x$, $y$, and $z$ (left to right). 
(d) Silicon nanopillar: magnitude of temperature gradient (left) and flux (right) for thermal loading along $x$, $y$, and $z$ (top to bottom).}
    \label{fig:Fig_sptc2fem_otherModels}
\end{figure}

We present analogous analyses for the Si twin grain boundary, polycrystalline nanowire, and nanopillar structures in Figure~\ref{fig:Fig_sptc2fem_otherModels}a--d. For the twin-grain-boundary system, the EMC-based mesh refinement produces a
finer mesh near the twin interface and along the outer surfaces in the $z$-direction (indicated by black arrows in
Figure~\ref{fig:Fig_sptc2fem_otherModels}a). These regions also exhibit larger temperature variations, whereas changes in $\|\mathbf{q}\|$ across the twin boundary are comparatively modest. This trend is consistent with Frequency-Domain Thermoreflectance (FDTR) measurements, which show that twin grain boundaries in Si are nearly transparent to heat flow, with only a small reduction in effective thermal conductivity~\cite{twinGB}, unless the interface is rough or nano-twinned (i.e., composed of closely spaced twin boundaries on the nanometer scale).

The polycrystalline structure illustrates this exception, as it contains both rough interfaces and nano-twinned regions. The slice shown in Figure~\ref{fig:Fig_sptc2fem_otherModels}b reveals pronounced spatial variation in both the directional SPTC field and the corresponding FE conductivity. We solve three heat-flow problems, one for each Cartesian direction, and in Figure~\ref{fig:Fig_sptc2fem_otherModels}c we show the magnitude of the temperature gradient $\|\nabla T\|$ (top) and the flux magnitude $\|\mathbf{q}\|$ (bottom) for thermal loading along $x$, $y$, and $z$ (from left to right). In all three load cases, both $\|\nabla T\|$ and $\|\mathbf{q}\|$ exhibit strong spatial variation, and the range of values differs between directions. Nevertheless, the effective conductivities obtained from homogenization, $K_\mathrm{eff}^\alpha$ (Equation~\ref{eq:K_eff_flux}), indicate that the structure is nearly isotropic, with $K_\mathrm{eff}^x = 14.99$, $K_\mathrm{eff}^y = 15.09$, and $K_\mathrm{eff}^z = 15.09~\mathrm{W\,m^{-1}\,K^{-1}}$. To obtain these values, we first rescale the SPTC-derived directional conductivity field by computing a volume-based effective conductivity $K_\mathrm{eff}^{\mathrm{vol},\alpha}$ and scaling $\mathbf{K}(\mathbf{x})$ to match a target conductivity of $15.15~\mathrm{W\,m^{-1}\,K^{-1}}$ (Equation~\ref{eq:K_scaled}), which corresponds to the SPTC-derived isotropic bulk conductivity (Equation~\ref{eq:SPTC_zeta}). This target value lies within the experimentally measured range for polycrystalline silicon thin films obtained using
time-domain thermoreflectance (TDTR)~\cite{Takeuchi2020PolySi}. 

Anisotropy is most pronounced in the nanopillar structure. In Figure~\ref{fig:Fig_sptc2fem_otherModels}d, we show $\|\nabla T\|$ (left) and $\|\mathbf{q}\|$ (right) for thermal loading along $x$, $y$, and $z$ (top to bottom). The masking and void-trimming procedure used to extract the pillars from the underlying brick mesh is described in Section~\textcolor{blue}{S2} (and Figure~\textcolor{blue}{S3}) of the supplemental material \cite{SM}. For thermal drives along the basal plane  ($x$ and $y$ directions),  the magnitudes of $\|\nabla T\|$ and $\|\mathbf{q}\|$ vary only weakly along the pillars. In contrast, loading along the $z$-direction, parallel to the pillars, produces much stronger spatial variations in both $\|\nabla T\|$ and $\|\mathbf{q}\|$, with the flux magnitude decreasing near the pillar–slab interface.  To quantify this anisotropy, we compute $K_\mathrm{eff}^\alpha$ for the three directions using the same volume-based scaling procedure as above (Equation~\ref{eq:K_scaled}), but with a target bulk conductivity of $27.98~\mathrm{W\,m^{-1}\,K^{-1}}$. The resulting flux-based effective conductivities are $K_\mathrm{eff}^x = 45.51$, $K_\mathrm{eff}^y = 45.37$, and $K_\mathrm{eff}^z = 13.20~\mathrm{W\,m^{-1}\,K^{-1}}$. The strong reduction in conductivity along the nanopillar axis is consistent with prior work on locally resonant nanophononic metamaterials for thermoelectric energy conversion~\cite{DavisNanopillars2014}.

\section{Experimental Validation of the Method}\label{subsec:Results_ExpVal}

\begin{table}[!t]
\centering
\caption{$2\mathrm{H}/3\mathrm{C}$ polytype silicon nanowire structural parameters. The experimental 2H-domain linear density ($n^{*}_{2\mathrm{H}}$) and 2H segment length ($L_{2\mathrm{H}}$) are taken from Reference~\cite{BenAmor2019}. Parentheses next to the experimental areal conductance $\Lambda_{\mathrm{exp}}$ denote the upper bound of the reported experimental uncertainty.}
\label{tab:polytype}
\begin{tabular}{ccccccccc}
\hline
Temp. [$^\circ$C] & $n^{*}_{2\mathrm{H}}$ [\textmu m$^{-1}$] & $L_{\mathrm{2H}}$ [\AA] & $\mathrm{s}~[\times 10^3] $ & $L^\mathrm{s}_{2\mathrm{H}}$ [\AA] & $L^\mathrm{s}_{3\mathrm{C}}$ [\AA] & d [\AA] & $\Lambda_\zeta$ [W/m$^2$K] & $\Lambda_\mathrm{exp}$ [W/m$^2$K]  \\
\hline
600  & 4.23 & 440 & 31 & 12.6 & 59.6 & 33.67 & 0.28 & 0.21 ($+$0.28) \\
700  & 5.59 & 460 & 39 & 18.9 & 53.3 & 33.84 & 0.23 & 0.12 ($+$0.14) \\
900  & 6.66  & 750 & 47 & 37.9 & 34.5 & 34.58 & 0.21 & 0.13 ($+$0.14)\\
1000 & 10.24 & 680 & 70 & 50.5 & 21.9 & 34.97 & 0.18 & 0.12 ($+$0.13) \\
\hline
\end{tabular}
\end{table}

We validate the effective conductivity predicted by \texttt{SCACS} against experimental conductivity measurements for cubic/hexagonal (3C/2H) silicon polytype nanowires with a diameter of 150~nm, produced via shear–stress–induced phase transformation at 600, 700, 900, and 1000~$^\circ$C, as reported by Ben~Amor
\textit{et~al.}~\cite{BenAmor2019,Vincent2018}. Following their work, we constructed four 3C/2H nanowire models using the reported 3C:2H length ratios, but scaling their overall dimensions for computational tractability. The model construction procedure is described in Section~\ref{subsubsec:models_expVal}, and the parameters used are summarized in Table~\ref{tab:polytype}.

\begin{figure}[!t]
    \centering
    \includegraphics[width=\linewidth]{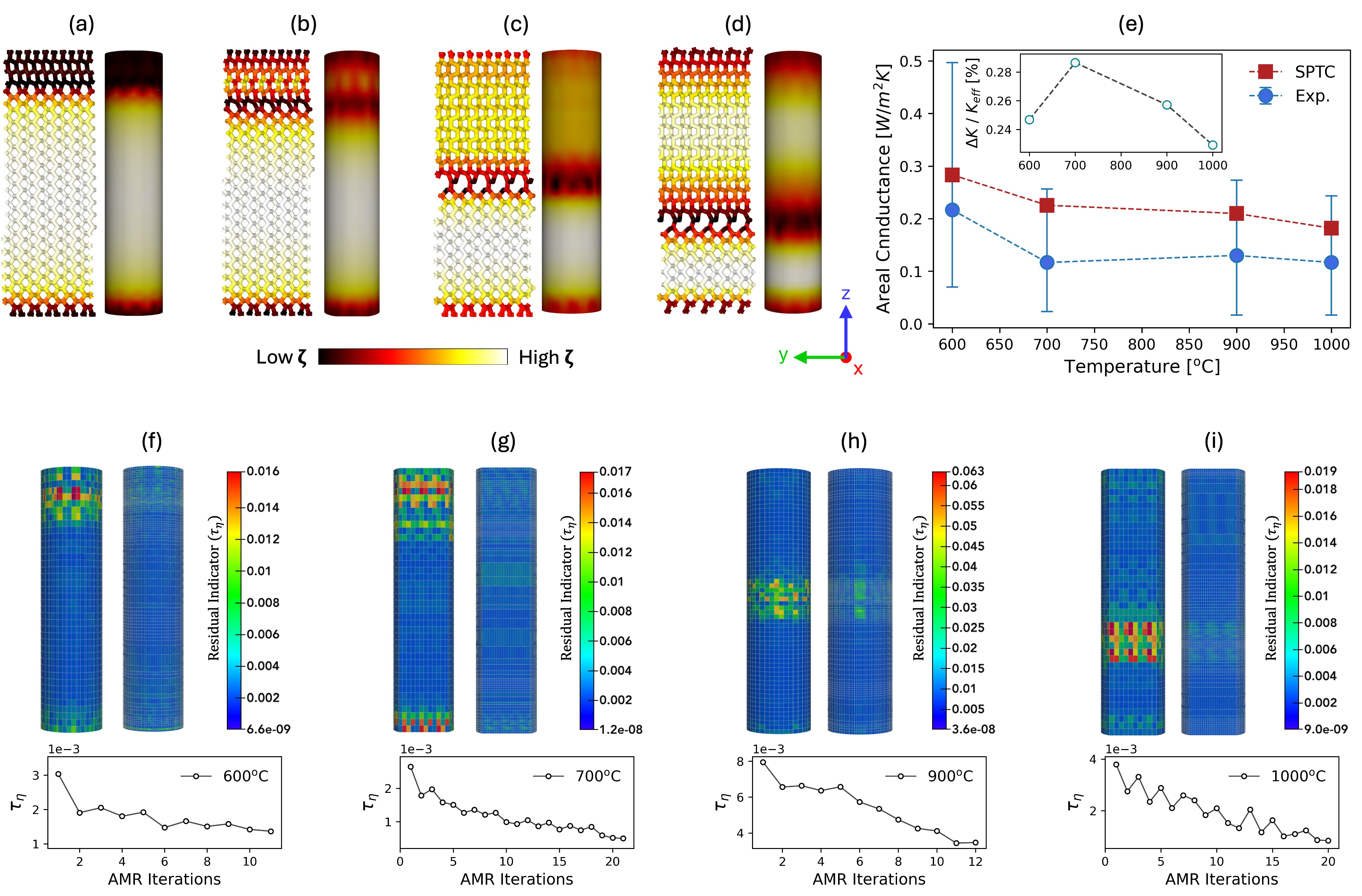}
    \caption{\textbf{Experimental validation of the method.} 
    (a--d) Atomic-scale (left) and FEM-scale (right) 3C/2H nanowire models at annealing temperatures of 600, 700, 900, and 1000~$^\circ$C. The colormap spans low (black) to high (white) SPTC.  (e) Experimental areal conductance~\cite{BenAmor2019} compared with SPTC-derived values (W\,m$^{-2}$\,K$^{-1}$); the inset shows the percent relative difference, $\Delta K = K_{\mathrm{exp}} - K_{\mathrm{eff}}$.  (f--i) Mesh refinement showing the starting coarse (left) and final refined (right) meshes. The bottom panel shows convergence of the residual indicator threshold $\tau_\eta$ for each refinement series.}
    \label{fig:Fig_sptc2fem_experiment}
\end{figure}

The isotropic SPTC distributions for the four polytype structures are shown in left panel of Figure~\ref{fig:Fig_sptc2fem_experiment}a--d, with the corresponding \texttt{sptc2fem} reconstructions on the right panel. In the 600~$^\circ$C model, the 3C region appears as the high-$\zeta$ (white) segment, whereas the 2H region appears in dark red (see Figure~\textcolor{blue}{S4} for the per-atom structural classification \cite{SM}). The SPTC contribution of the 2H segment increases with its scaled length $L_{2\mathrm{H}}$ as the growth temperature rises, reflecting a transition from a small defect-like perturbation at 600~$^\circ$C to a more substantial heterostructure at 900–1000~$^\circ$C. Across all four polytype models, the 3C segment remains the dominant contributor to the total conductivity.

Thermal conductance of Si nanowires depends strongly on diameter~\cite{Doerk2010}, and our simulated wires have significantly smaller diameters than the experimental ones. To enable a consistent comparison, we introduce a simple \emph{diameter-normalized areal conductance} metric, \( \Lambda = \frac{K_\mathrm{eff}}{d}, \) i.e., the effective axial thermal conductivity divided by the nanowire diameter (listed as $d$ in Table~\ref{tab:polytype}). This rescaling compensates for the difference between our modeled diameters and the experimental value of 150~nm. As summarized in Table~\ref{tab:polytype} and shown in Figure~\ref{fig:Fig_sptc2fem_experiment}e, the \texttt{SCACS}-derived areal conductance $\Lambda_{\zeta}$ (red) follows the same temperature trend as the experimental
areal conductance $\Lambda_{\mathrm{exp}}$ (blue) and lies within the upper part of the reported experimental uncertainty range. This supports the use of our SPTC-based atomistic models as representative volume elements for continuum
modeling of the 3C/2H Si polytype material.

Treating the FEM models as RVEs, we compute the effective conductivity along the
$z$-direction, $K_\mathrm{eff}^z$, which is normal to the 3C/2H interface. As in the polycrystalline case, each 3C/2H model is first scaled to match the corresponding experimental average conductivity (used as the target), with the \texttt{SCACS}-based $\mathbf{K}(\mathbf{x})$ acting as a spatial weighting field for the heterogeneous response. The experimental conductivities were approximately 32.5, 17.5, 19.5, and 17.5~$\mathrm{W\,m^{-1}\,K^{-1}}$ for nanowires annealed at 600, 700, 900, and 1000~$^\circ$C, respectively. As shown in the inset of Figure~\ref{fig:Fig_sptc2fem_experiment}e, the relative differences between these experimental values and the \texttt{SCACS} flux-based effective conductivities remain small (0.23--0.26\%), with $K_\mathrm{eff}^z \approx 32.42$, 17.46, 19.45, and 17.46~$\mathrm{W\,m^{-1}\,K^{-1}}$ at 600, 700, 900, and 1000~$^\circ$C, respectively. This rescaling ensures that each FE model reproduces the measured bulk response while preserving the internal 3C/2H contrast, supporting their use as RVEs for the thermal behavior of 3C/2H silicon heterostructures.

Figures~\ref{fig:Fig_sptc2fem_experiment}f--i show the coarse (left) and refined (right) meshes generated by our EMC-based AMR procedure. A starting grid of $20\times 20\times 40$ elements provides a good compromise between accuracy and
computational cost. The residual indicator $\eta$, plotted as a colormap from blue (low) to red (high), clearly shows that refinement is required near the 3C/2H interfaces for all models. Convergence is reasonably fast, and the final meshes—
plotted using a common color scale for both coarse and refined discretizations— exhibit an essentially uniform residual, as expected for a well-refined mesh~\cite{zienkiewicz2005finite}.

For the four models considered, the optimal meshes contain 40, 50, 38, and 46 elements in the $x$--$y$ plane and 80, 74, 82, and 70 elements in the thermal loading (\(z\)) direction at 600, 700, 900, and 1000~$^\circ$C, respectively. As
indicated by the white grid-lines in Figure~\ref{fig:Fig_sptc2fem_experiment}f--i, our implementation of the flux-aware effective element size $h_\mathcal{K}^*$ enforces refinement in boundary regions of highest flux, thereby reducing the need to manually manipulate the model geometry to expose interfaces, as is often done in conventional FE studies.

\section{Constructing Orthotropic Conductivity Tensors from Directional Data}
\label{Results_subsec:orthotropic_from_directional}

In many applications, thermal transport is first characterized by directional conductivities along three mutually orthogonal directions. These may coincide with the global Cartesian axes $(x,y,z)$ or with a rotated set of material axes (e.g.\ grain- or interface-aligned directions). In either case, the goal is to construct an orthotropic effective thermal conductivity tensor $\mathbf{K}_{\mathrm{eff}}$ that can be used in continuum solvers. For a given RVE, \texttt{SCACS} computes effective conductivities for unit temperature gradients applied along the global axes, yielding \( K^x_{\mathrm{eff}},\,K^y_{\mathrm{eff}},\,K^z_{\mathrm{eff}}. \) If the material principal axes coincide with $(x,y,z)$, the orthotropic tensor in global coordinates is simply \(  \mathbf{K}_{\mathrm{eff}}^{\mathrm{global}} = \mathrm{diag}\bigl(
K^x_{\mathrm{eff}},\,K^y_{\mathrm{eff}},\,K^z_{\mathrm{eff}}\bigr)\). 

More generally, the principal directions of heat conduction need not align with the global axes. Suppose the effective conductivities $K^1_{\mathrm{eff}},K^2_{\mathrm{eff}},K^3_{\mathrm{eff}}$ act along three orthonormal material axes represented by unit vectors
\(
\{\mathbf{d}_1,\ \mathbf{d}_2,\ \mathbf{d}_3\}.
\) In this local material basis, the effective conductivity tensor is diagonal: \( \mathbf{K}^{\mathrm{local}}_{\mathrm{eff}} = \mathrm{diag}\bigl(K^1_{\mathrm{eff}},K^2_{\mathrm{eff}},K^3_{\mathrm{eff}}\bigr).
\) If $V$ denotes the $3\times 3$ orientation matrix whose columns are the material axes expressed in the global frame, i.e.\ $V = [\,\mathbf{d}_1\
\mathbf{d}_2\ \mathbf{d}_3\,]$. The global orthotropic conductivity tensor then follows from a standard change of basis:
\begin{equation}
  \mathbf{K}^{\mathrm{global}}_{\mathrm{eff}}
  = V\,\mathbf{K}^{\mathrm{local}}_{\mathrm{eff}}\,V^\mathsf{T}
  = \sum_{\beta=1}^3 K^\beta_{\mathrm{eff}}\,\mathbf{d}_\beta \otimes \mathbf{d}_\beta .
\end{equation}
The resulting $\mathbf{K}^{\mathrm{global}}_{\mathrm{eff}}$ is symmetric and positive semi-definite provided $K^\beta_{\mathrm{eff}} \ge 0$, and can be supplied directly to continuum solvers as the full anisotropic conductivity tensor.

In practice, this construction allows users who may prefer other finite-element or continuum solvers (e.g.\ Abaqus, COMSOL, ANSYS, or computational fluid dynamics (CFD) codes) to import conductivity tensors obtained from \texttt{SCACS} without explicitly building atomistic models, provided they specify the principal conductivities $(K^1_{\mathrm{eff}},K^2_{\mathrm{eff}},K^3_{\mathrm{eff}})$ together with a material orientation defined by $V$ (or, equivalently, by Euler angles or direction cosines). In the special case where the material axes coincide with the global axes, $V=\mathbf{I}$ and the global tensor reduces to the diagonal form above.

\section{Discussion}

\subsection{Structured and Unstructured Mesh Implementation}

All results in this work are obtained on a structured ``brick" hexahedral mesh. Nevertheless, it is straightforward to reconstruct the SPTC field on an unstructured mesh by interpolating the brick-wise nodal tensor field $\mathbf{K}_n$; we denote the resulting tensor field by $\mathbf{K}'_n$:
\begin{equation}
    \mathbf{K}'_n \approx \mathcal{I}_h[\mathbf{K}_n]
\end{equation}
where $\mathcal{I}_h$ denotes a suitable interpolation operator.

An unstructured mesh can offer improved geometric fidelity for irregular cross-sections and finer control over element size near boundaries for mesh-enrichment procedures, although EMC already performs satisfactorily on the hexahedral mesh (see Section~\ref{subsubsec:amr-sptc}). However, for unstructured meshes, particular care must be taken when performing the coordinate transformation into the SPTC frame so as to control numerical errors and to preserve the symmetry and positive semi-definiteness of $\mathbf{K}'_n$.

\subsection{Role of the Conductivity Tensor in the Energy Norm and Error Measures}

The bilinear form of the FE formulation (LHS of Equation \ref{eq:Continous_WeakForm}) induces an energy norm, which is the
natural norm for both the continuous and discrete problems
\cite{zienkiewicz2005finite}. If $T$ denotes the exact solution and $T^h$ its
FE approximation, we define the nodal error vector $\varepsilon = T - T^h$. The global
discrete energy norm of the error is then
\begin{equation}
  \|\varepsilon\|_{E,K}^2 = \varepsilon^\mathsf{T} \mathbf{K}\, \varepsilon
\end{equation}
where the global stiffness matrix (conductivity tensor $\mathbf{K}$) is assembled from the element matrices:
\begin{equation}
  [\mathbf{K}^e]_{ab}
  = \int_{\Omega^e} \mathbf{G}_a^\mathsf{T}\,\mathbf{k}^e\,\mathbf{G}_b\,\mathrm{d}\Omega \, ;
  \qquad
  \mathbf{K} = \sum_e \mathbf{K}^e
\end{equation}
as in the first term of Equation~\ref{eq:FE_WeakForm}. Thus, the squared
energy norm of the error is exactly the difference between the energies of the exact and FE solutions, measured with the same conductivity tensor $\mathbf{K}$.

For an isotropic material with true scalar conductivity $k$ such that
$\mathbf{K} \approx k\mathbf{I}$, and an idealized or extrapolated scalar value
$k^*$, the physical interpretation is straightforward: regions with high (or low)
true conductivity are systematically underestimated if $k^* < k$ and
overestimated if $k^* > k$. This distortion becomes more severe in anisotropic
materials when oversimplified conductivity tensors are used and directional
gradients are mis-weighted. In that case, the FE solution effectively solves
the governing equation for a surrogate conductivity field, and the associated
energy norm and error estimates quantify accuracy with respect to this
surrogate model, rather than the true material.

In our approach, the SPTC-based conductivity tensor is applied at the element
level and represents an RVE of the material. Anisotropy is therefore correctly encoded in how the energy norm
penalizes gradients and fluxes, without relying on \textit{ad hoc} geometric tuning or
mesh alignment, and the FE solution energy $\|T^h\|_E$ provides a more faithful
approximation to the physical energy $\|T\|_E$.

\subsection{Thermodynamic and Energy Consistency of the Conductivity Tensor}

The applicability of the SPTC$\to$FEM mapping rests on local thermodynamic equilibrium, i.e., the existence of an intermediate coarse-graining length $\ell_{\text{vox}}$ over which intensive variables (such as the temperature $T$) are well defined and gradients are smooth, while $\ell_{\text{vox}}$ remains much larger than the atomic spacing $a$, such that:
\begin{equation}
  a \ll \ell_{\text{vox}} \ll \ell_T
\end{equation}
This scale separation criterion assumes diffusive transport (Fourier’s law) and requires that the characteristic length scale $\ell_T$ over which $T$ varies macroscopically is much larger than the phonon mean free path, $\lambda \ll \ell_T$. For silicon, $\lambda \approx 8\,\text{\AA}$ \cite{Ugwumadu2025} is significantly smaller than the cell size of the dataset used to build \texttt{sptc-ai}, so the voxel-wise conductivities inferred from atomistic data are physically meaningful inputs to the continuum PDE.

Additionally, the significance of our method lies not only in its transparent physical interpretation, but also in the fact that it admits clear, independent consistency checks. The conductivity field obeys the Hill--Mandel energy consistency condition~\cite{Hill1972O,mandel1972,Roberto2017}, which equates macroscopic and microscopic power:
\begin{equation}
  \nabla T_{\mathrm{m}}^\mathsf{T} \; \mathbf{K}_\mathrm{eff}\,\nabla T_{\mathrm{m}}
  =
  \frac{1}{|\Omega|}\int_\Omega \nabla T^\mathsf{T} \;\mathbf{K}(\mathbf{x})\,\nabla T\,\mathrm{d}\mathbf{x}
\end{equation}
and, for a uniform macroscopic gradient $\nabla T_{\mathrm{m}} = \nabla T$, this yields an effective heat flux. When $\mathbf{K}(\mathbf{x})$ is constructed from SPTC, this homogenized tensor $\mathbf{K}_\mathrm{eff}$ reproduces the Allen--Feldman bulk conductivity by design (up to voxelization and quadrature errors). From the atomistic side, the Allen--Feldman conductivity decomposition guarantees linear additivity of modal contributions to thermal transport~\cite{Ugwumadu2025}, so any voxelized tensor field $\mathbf{K}(\mathbf{x})$ assembled from $\{\zeta_i^\alpha\}$ recovers the correct bulk response in the homogeneous limit.

As an outlook for future work, the availability of full temperature and heat-flux fields makes experimental validation direct. Once $\mathbf{K}(\mathbf{x})$ is specified, the simulated temperature and heat flux can be computed under the same boundary conditions as in experiment (see Section~\ref{subsec:Results_ExpVal}). The \texttt{sptc2fem} mapping enriches the spatial detail of $\mathbf{K}(\mathbf{x})$ by embedding local atomic-scale variations into the continuum field, preserving anisotropy and inhomogeneity inherited from molecular or \textit{ab initio} data. As a result, the temperature field captures local thermal bottlenecks and directional effects that are inaccessible to traditional uniform-conductivity FEM models.

\section*{Data Availability}
The data that support the findings of this article are openly available \cite{Zenodo}. The source code for \texttt{SCACS} will be available upon request subject to LANL and US export control policies. 

\begin{acknowledgments}
The authors thank Dr. Jianxin Zhu, Dr. Gian Luca Delzanno, Dr. Pedro Alberto Resendiz Lira, and Dr. Christopher Lane (all at Los Alamos National Laboratory) for discussions that greatly contributed to this work.
C.U. and R.M.T. acknowledge support from the Laboratory Directed Research and Development (LDRD) program at Los Alamos National Laboratory (LANL) through the Director’s Postdoctoral Fellowship Program (20240877PRD4) and the Exploratory Research (ER) program (20240397ER). LANL is operated by Triad National Security, LLC, for the U.S. Department of Energy (DOE) National Nuclear Security Administration under Contract No. 89233218CNA000001. 
\end{acknowledgments}

\section*{Author Contributions}
C.U. conceived the idea and wrote the manuscript. D.A.D. and R.M.T contributed to revision of the manuscript. All authors discussed the results, implications and commented on the manuscript at all stages. 

\appendix 

\section{Silicon Structures: Construction and Optimization}\label{subsec:models}

\subsection{Training, Validation, and Test Structures}\label{subsubsec:models_training}
The \texttt{sptc-ai} graph neural network (GNN) model was trained on a corpus of 653 silicon structures, including 607 models obtained from Rosset \textit{et al.}~\cite{rosset2025}, supplemented by additional configurations generated in this work. The Rosset dataset (\emph{R-models}) consists of 1000-atom amorphous, paracrystalline, and polycrystalline networks. To expand structural diversity, we added two complementary sets containing between 998–1356 atoms each: (i) 23 polycrystalline cells with varied grain boundaries (\emph{P-models}), and (ii) 23 crystalline cells (\emph{Q-models}) representing the cubic diamond phase (3C-Si) and the hexagonal wurtzite phase (2H-Si), some of which include vacancy or interstitial defects. All additional structures constructed in this work were generated using \texttt{Atomsk}~\cite{ATOMSK}. Visualization of the atomic structure was done using OVITO \cite{OVITO}

\subsection{Model Evaluation and Finite-element Structures}\label{subsubsec:models_fea}
To assess the generalization performance of \texttt{sptc-ai}, we included a 2022-atom amorphous–crystalline (AC) Si interface model whose SPTC characteristics were previously reported~\cite{Ugwumadu2025}. For the finite-element implementation (and in addition to the AC model), we constructed three large representative volume element (RVE) models: a 223{,}930-atom Si twin-grain-boundary nanowire, a 134{,}923-atom polycrystalline Si nanowire, and a 26{,}191-atom Si nanopillar.

\subsection{Experimental Validation Structures}\label{subsubsec:models_expVal}
Four additional nanowire models were generated to represent the 3C/2H (cubic/hexagonal) polytype structures experimentally fabricated and studied by Ben Amor \textit{et al.}~\cite{BenAmor2019}. The 3C/2H nanostructures were obtained at annealing temperatures of 600, 700, 900, and 1000~$^\circ$C. The lattice constants were set to $a_0 = 5.431$~\AA{} for 3C-Si and $a_{\text{hex}} = 3.84$~\AA{}, $c_{\text{hex}} = 6.31$~\AA{} for 2H-Si. Following Reference~\cite{BenAmor2019}, the crystallographic orientations were 3C-Si: $x \parallel [1\bar{1}0]$, $y \parallel [001]$, $z \parallel [110]$,  and 2H-Si: $x \parallel [\bar{2}110]$, $y \parallel [01\bar{1}0]$, $z \parallel [0001]$, with the interface normal along $z$.  

The heterostructure geometry was determined from the experimental 2H segment length ($L_{2\mathrm{H}}$) and 2H-domain linear density ($n^{*}_{2\mathrm{H}}$) from Reference \cite{BenAmor2019}. Assuming a single 2H domain per cell, we estimate the scaled 3C segment length, $L^{s}_{3\mathrm{C}}$ as:
\begin{equation}\label{eq:polytype}
    L^{s}_{3\mathrm{C}} = \frac{10^4\, s}{n^{*}_{2\mathrm{H}}} - L^{s}_{2\mathrm{H}}
\end{equation}
where $L^{s}_{2\mathrm{H}} = s \, L_{2\mathrm{H}}$, and $s$ is a scaling factor used to reduce overall system size while preserving the experimental 2H:3C ratio, yielding a uniform system size of approximately 2300 atoms per structure. The factor $10^4$ converts $n^{*}_{2\mathrm{H}}$ from $\mu\mathrm{m}^{-1}$ to \AA$^{-1}$. The values of $L_{2\mathrm{H}}$, $s$, and $n^{*}_{2\mathrm{H}}$ are summarized in Table~\ref{tab:polytype}.

\subsection{Atomic Structure Optimization Protocol}\label{subsec:models_relaxation}
All the atomic-scale models were relaxed following a multi-stage conjugate-gradient (CG) energy minimization procedure using the machine-learning Gaussian Approximation Potential (GAP) for silicon~\cite{Deringer2021}, as implemented in the Large-scale Atomic/Molecular Massively Parallel Simulator (LAMMPS) package~\cite{lammps}. Atomic positions and cell shape were allowed to change under zero external pressure, with a maximum strain increment of 0.001 per iteration to avoid unphysical volume changes. A quadratic line search was used with a maximum atomic displacement of 0.1~\AA{} per iteration, and convergence was monitored using the maximum force norm. Relaxation proceeded in three stages:  (1) a coarse stage with a force tolerance of $10^{-4}$~eV/\AA{},  (2) a refinement stage with tolerance $5\times 10^{-6}$~eV/\AA{}, and  (3) a final tightening stage with tolerance $10^{-6}$~eV/\AA{}.

\section{\texttt{sptc-ai} Architecture}\label{subsec:sptc_ai}

\subsection{Node, Edge, and Global Features}\label{subsubsec:features}

Each atom (node) is represented by a 91-dimensional feature vector comprising 64 principal components of the Smooth Overlap of Atomic Positions (SOAP) descriptor~\cite{SOAP} and an additional 27-dimensional block. The latter includes: (i) structural descriptors such as one-hot structural phase encoding, phase confidence, grain-boundary flag, and vacancy-cluster size; (ii) OVITO-based descriptors \cite{OVITO} encoding the Si structure class and coordination-shell index (CSI) depth; and (iii) statistical descriptors capturing local coordination number (\texttt{CN}), first-shell angular distortion (\texttt{$\angle$rms}), bond-length statistics,  and Steinhardt order parameters ($q_4$, $q_6$). A description for the node features is provided in Tables~\textcolor{blue}{S1} and \textcolor{blue}{S2} \cite{SM}.

The statistical descriptors are central to curating the training set.  The coordination number for each atom \(i\) is defined as the number of neighbors \(j\) satisfying \(d_{ij}\le2.80~\text{\AA}\). The first-shell angular distortion quantifies deviation from ideal tetrahedral geometry as,
\(
\texttt{$\angle$rms}_i = \sqrt{\tfrac{1}{6}\sum_{i<j}(\theta_{ij} - \theta^*)^2},
\) where \(\theta_{ij}\) are the six inter-bond angles among the four nearest neighbors and \(\theta^* = \idealTetra\) is the ideal tetrahedral angle. Bond-length statistics are evaluated within the same cutoff of $2.80~\text{\AA}$, including the mean, \( B_{\mu,i} = \tfrac{1}{n_i}\sum_{d\in\mathcal{B}_i} d, \) and standard deviation, \( B_{\sigma,i} = \sqrt{\tfrac{1}{n_i}\sum_{d\in\mathcal{B}_i}(d - B_{\mu,i})^2}, \) where \(n_i = |\mathcal{B}_i|\). From the bond-length tails, the short- and long-bond fractions (\texttt{B$_\text{S}$} and \texttt{B$_\text{L}$}) are defined as
\begin{align}
B_{\text{S},i} &= \frac{1}{n_i}\sum_{d\in\mathcal{B}_i}\mathbf{1}\{d < \tau_i^-\} \\
B_{\text{L},i} &= \frac{1}{n_i}\sum_{d\in\mathcal{B}_i}\mathbf{1}\{d > \tau_i^+\}
\end{align}
with thresholds \(\tau_i^\pm = B_{\mu,i} \pm 0.05~\text{\AA}\).  

Each directed edge $(j\!\to\!i)$ carries a 35-D attribute vector composed of a 32-channel Gaussian-type radial basis function (RBF) expansion of the interatomic distance and a three-component unit displacement under the minimum-image convention (MIC):
\begin{align}\label{eq:edgeFeatures}
\mathbf{r}_{ij}
&=
\Bigl[\underbrace{\phi_1(d_{ij}),\ldots,\phi_{32}(d_{ij})}_{\text{RBF (32-D)}}
\;\big\|\;
\underbrace{\hat{u}_x,\hat{u}_y,\hat{u}_z}_{\text{unit MIC direction}}
\Bigr]
\\
\hat{\mathbf{u}}&=\frac{\mathrm{MIC}(\mathbf{r}_j-\mathbf{r}_i)}{\|\mathrm{MIC}(\mathbf{r}_j-\mathbf{r}_i)\|_2}
\end{align}

The expansion centers, widths, and the cutoff radius $r_{\mathrm{cut}}=5.5~\text{\AA}$ are fixed during dataset generation. This value is chose to match the optimal cutoff identified for SPTC analysis in silicon \cite{Ugwumadu2025}. Edge attributes are computed once during featurization and serialized together with raw neighbor distances for verification. 

At the (global) graph level, a 21-dimensional global feature vector $\mathbf{g}_g$ is constructed for use in the Feature-wise Linear Modulation (FiLM)~\cite{perez2018film} implementation of the graph network model. This vector combines multi-scale structural statistics, including the mean and standard deviation of Steinhardt order parameters ($q_4$, $q_6$), coordination number, and vacancy-cluster size; the global crystalline fraction; and a four-bin histogram (fractional count) of CSI depths ($h^{\mathrm{CSI}}_{0-3}$). It further includes the simulation box lengths $(L_x,L_y,L_z)$, and a four-category regime one-hot encoding (crystal, polycrystalline, paracrystalline, amorphous) with its confidence score. Missing global fields are zero-filled to preserve a fixed length of $d_{\mathrm{glob}}=21$. Scalar descriptors are standardized (z-scored) using statistics from the training split, while categorical and embedding features remain unscaled. Additional details on the global node features and the regime classification protocol  are provided in Section \textcolor{blue}{S1.2} and \textcolor{blue}{S1.3}, respectively \cite{SM}.

\subsection{The Graph Network Model}\label{subsubsec:sptc_ai_gin}
The \texttt{sptc-ai} model is a message-passing graph neural network with $L=4$ layers and hidden node dimension $d_h = 256$. Each layer is a GIN-style block (but without an explicit $(1+\varepsilon)$ self-weight) that updates node states using neighbor aggregation, multi-layer perceptron (MLP) transformations, residual connections, FiLM modulation, and layer normalization (LN). 

If $h_i^{(\ell)} \in \mathbb{R}^{d_h}$ denotes the hidden state of node $i$ at layer $\ell$, and  $\mathcal{N}(i)$ be the set of neighbors of node $i$.
For an edge $(i,j)$ we denote by $e'_{ij} \in \mathbb{R}^{d_h}$ the edge-attribute embedding. Within layer $\ell$, the updates are:
\begin{align}
s_i^{(\ell)} &=
\sum_{j\in\mathcal{N}(i)} \rho\!\big(h_j^{(\ell-1)} + e'_{ij}\big)
\label{eq:aggregator}\\
z_i^{(\ell)} &=
\phi_g\!\big(h_i^{(\ell-1)} + s_i^{(\ell)}\big)
\label{eq:nodeMLP}\\
\tilde{h}_i^{(\ell)} &=
h_i^{(\ell-1)} + z_i^{(\ell)}
\label{eq:preNormRes}\\
\widetilde{h}_i^{(\ell)} &=
\tilde{h}_i^{(\ell)} \odot \bigl(1 + \boldsymbol{\gamma}_{b(i)}\bigr)
+ \boldsymbol{\beta}_{b(i)}
\label{eq:FiLMmodulation}\\
\hat{h}_i^{(\ell)} &= \mathrm{LN}\!\big(\widetilde{h}_i^{(\ell)}\big)
\label{eq:LN}\\
h_i^{(\ell)} &=
\hat{h}_i^{(\ell)} + \phi_f\!\big(\hat{h}_i^{(\ell)}\big)
\label{residualFFN}
\end{align}
where $\rho(u)=\max(0,u)$ is a ReLU nonlinearity applied in the aggregation step (Equation~\ref{eq:aggregator}). The index $b(i)$ maps node $i$ to its graph in the mini-batch (i.e.\ $b(i)$ is the graph ID), so that $\boldsymbol{\gamma}_{b(i)},\boldsymbol{\beta}_{b(i)} \in \mathbb{R}^{d_h}$
are graph-level FiLM parameters shared by all nodes in the same graph.

The node-wise MLPs $\phi_g$ and $\phi_f$ are both of the generic form
\begin{equation}
  \phi_k(u) = W_{k,2}\,\sigma\!\big(W_{k,1}\,u + b_{k,1}\big) + b_{k,2}
\end{equation}
where $k\in\{g,f\}$, $W_{k,1},W_{k,2}$ and $b_{k,1},b_{k,2}$ are learned weights and biases, and $\sigma$ is a GELU activation function~\cite{GELUpaper}. Thus, ReLU is used only in the aggregation step, while all MLPs use GELU nonlinearities. The FiLM adapter in Equation~\ref{eq:FiLMmodulation} applies per-graph scale and shift parameters $(\boldsymbol{\gamma}_g,\boldsymbol{\beta}_g)$ constructed from a global descriptor $\mathbf{g}_g$ of graph $g$ via a small conditioning network; for a node $i$ in graph $g$, we write $\boldsymbol{\gamma}_{b(i)} = \boldsymbol{\gamma}_g$ and $\boldsymbol{\beta}_{b(i)} = \boldsymbol{\beta}_g$.

After the final message-passing layer, the node states are passed through a final layer normalization:
\(
  \bar{h}_i = \mathrm{LN}\big(h_i^{(L)}\big).
\) These normalized node embeddings are then fed into two linear output heads, one predicting the three Cartesian components of the SPTC vector and the other its isotropic value, given as:
\begin{align}\label{eq:GNNheads}
\hat{\zeta}^\mathbf{r}_i &=
W_{\mathbf{r}}\,\bar{h}_i + b_{\mathbf{r}} \in \mathbb{R}^3
\\[3pt]
\hat{\zeta}^\mathrm{d}_i &=
W_d\,\bar{h}_i + b_d \in \mathbb{R}
\end{align}
where $W_{\mathbf{r}} \in \mathbb{R}^{3\times d_h}$, $b_{\mathbf{r}}\in\mathbb{R}^3$, $W_d\in\mathbb{R}^{1\times d_h}$, and $b_d\in\mathbb{R}$ are learned parameters. Since SPTC is defined per atom, no global pooling is applied.

In addition, we define a ``composite'' isotropic prediction obtained as the simple arithmetic mean of the three directional components:
\begin{equation}\label{eq:compHead}
  \hat{\zeta}^\mathrm{c}_i =
  \tfrac{1}{3}\bigl(\hat{\zeta}^x_i + \hat{\zeta}^y_i + \hat{\zeta}^z_i\bigr)
\end{equation}
which can be used as an auxiliary isotropic proxy or consistency check between the vector-valued and scalar SPTC predictions.

\subsection{Training, Metrics, and Evaluation}\label{subsubsec:sptc_ai_trainMetricEval}

Training was performed using the AdamW optimizer~\cite{loshchilov2019} with learning rate $5\times 10^{-6}$, weight decay $10^{-2}$, mini-batch size $|B|=8$, and early stopping based on validation loss with a patience of
20~epochs. Gradients were clipped via $\|\nabla_\theta \mathcal{L}_{\mathrm{t}}\|_2 \le 5$ for numerical stability. With these settings, convergence was reached by epoch~83 within a 400-epoch budget.

All losses were computed in standardized target space, i.e.\ each target component was normalized using dataset-level mean and standard deviation before computing the loss, and metrics were reported after de-standardization. For each mini-batch $B$ of graphs, per-node predictions $\{\hat{\boldsymbol{\zeta}}_i\} = \hat{\zeta}^x_i,\hat{\zeta}^y_i,\hat{\zeta}^z_i,\hat{\zeta}^\mathrm{d}_i)$
were compared against targets $\{\boldsymbol{\zeta}_i\} =
(\zeta^x_i,\zeta^y_i,\zeta^z_i,\zeta^\mathrm{d}_i)$. The total loss $\mathcal{L}_{\mathrm{t}}$ was defined as
\begin{align}
\mathcal{L}_{\mathrm{t}} &=
\mathcal{L}_{\mathrm{\alpha}} + \mathcal{L}_{\mathrm{c}}
+ \mathcal{L}_{\mathrm{d}}\\[3pt]
\mathcal{L}_{\mathrm{\alpha}}
&=\mathbb{E}_B\!\big[(\hat{\zeta}^x_i-\zeta^x_i)^2\big]
 +\mathbb{E}_B\!\big[(\hat{\zeta}^y_i-\zeta^y_i)^2\big]
 +\mathbb{E}_B\!\big[(\hat{\zeta}^z_i-\zeta^z_i)^2\big]\\[3pt]
 \mathcal{L}_{\mathrm{c}}
&=\mathbb{E}_B\!\big[(\hat{\zeta}^{\mathrm{c}}_i - \zeta^\mathrm{d}_i)^2\big]\\[3pt]
\mathcal{L}_{\mathrm{d}}
&=\mathbb{E}_B\!\big[(\hat{\zeta}^{\mathrm{d}}_i - \zeta^\mathrm{d}_i)^2\big]
\;+\;\beta\mathcal{L}_{\mathrm{c}}
\end{align}

\noindent where $\mathcal{L}_{\mathrm{\alpha}}$, $\mathcal{L}_{\mathrm{c}}$, and $\mathcal{L}_{\mathrm{d}}$ are the directional, consistency, and isotropic loss, respectively, with $\beta=0.1$.  The consistency loss $\mathcal{L}_{\mathrm{c}}$ uses the composite isotropic prediction $\hat{\zeta}^\mathrm{c}_i$ defined in Equation~\ref{eq:compHead}. For a mini-batch $B$ consisting of graphs $g$ with vertex sets $V^{(g)}$, the batch expectation operator $\mathbb{E}_B[\cdot]$ is defined by:
\begin{equation}\label{eq:batchExpectation}
\mathbb{E}_B[q]
= \frac{1}{|B|}
\sum_{g\in B} \frac{1}{|V^{(g)}|}
\sum_{i\in V^{(g)}} q_i
\end{equation}
i.e.\ a uniform average over all nodes in all graphs in the batch.

Model performance was assessed using  mean-absolute error (MAE) and root-mean-square error (RMSE):
\begin{equation}\label{eq:MAE_RMSE}
\mathrm{MAE}_t =\tfrac{1}{N}\sum_i|\hat{\zeta}^t_i-\zeta^t_i|\,; \qquad
\mathrm{RMSE}_t=\sqrt{\tfrac{1}{N}\sum_i(\hat{\zeta}^t_i-\zeta^t_i)^2}
\end{equation}
for $t\in\{x,y,z,\mathrm{d}\}$,  where $N$ is the total number of nodes in the evaluation set. Performance was evaluated both globally and within the OVITO-based structural classes ($k\in\{0,\dots,6\}$). Prediction consistency
between the scalar and vector heads was quantified via:
\begin{equation}
\Delta_i
= \Big|\hat{\zeta}^\mathrm{d}_i
- \tfrac{1}{3}\big(\hat{\zeta}^x_i+\hat{\zeta}^y_i+\hat{\zeta}^z_i\big)\Big|
\end{equation}
and summarized using the mean, median, $95^\text{th}$ percentile, and maximum of $\{\Delta_i\}$ over the evaluation set.

\section{\texttt{sptc2fem} Architecture}\label{subsec:sptc2fem}

\subsection{The Finite Element Equation and  Solve Procedure}\label{subsubsec:heatEquation}

From the continuity equation (Equation \ref{eq:continuityRelation}) and constitutive relation (Equation \ref{eq:Fourier}) for the heat transport probelm, we define the computational domain $\Omega\subset\mathbb R^3$, and the boundary
\(
\partial\Omega = \Gamma_D \,\dot\cup\, \Gamma_N \,\dot\cup\, \Gamma_R ,
\) where $\Gamma_D$, $\Gamma_N$, and $\Gamma_R$ correspond to Dirichlet, Neumann, and Robin boundaries, respectively. The prescribed boundary conditions with a volumetric source $f(\mathbf x)$ are:

\begin{align}
 \nabla^\mathsf{T} \big(\mathbf K(\mathbf x)\nabla T \big) &= f(\mathbf x)  && \text{in } \Omega \label{eq:strongForm}\\
 T &= \bar T && \text{on } \Gamma_D\\
 q_n  &= \bar q && \text{on } \Gamma_N\\
  h\,(T - T_\infty) &= -q_n && \text{on } \Gamma_R
\end{align}
where $\bar T$ is the prescribed temperature, $h$ is the heat transfer or radiation coefficient, $T_\infty$ is the ambient temperature, $\bar q$ is a specified flux value, and \(
    q_n = \mathbf n^\mathsf{T} \big(\mathbf K(\mathbf x)\nabla T\big),
\) with $\mathbf n$ the outward unit normal on the boundary.

The weak formulation of Equation \ref{eq:strongForm} is given as \cite{zienkiewicz2005finite}:
\begin{equation}\label{eq:Continous_WeakForm}
\int_{\Omega} \nabla v^\mathsf{T} \bigl(\mathbf{K}(\mathbf{x}) \nabla T\bigr)\,\mathrm{d}\Omega
\;+\; \int_{\Gamma_R} h\,v\,T\,\mathrm{d}\Gamma
= -\int_{\Omega} v\,f(\mathbf{x})\,\mathrm{d}\Omega
\;+\; \int_{\Gamma_N}  v\,\bar{q}\,\mathrm{d}\Gamma
\;+\; \int_{\Gamma_R} h\,T_{\infty}\,v\;\mathrm{d}\Gamma
\end{equation}
valid for all test functions $v \in \mathcal{V} := \{v \in H^1(\Omega)\,|\, v = 0 \text{ on }\Gamma_D\}$, where $H^1(\Omega)$ is the Sobolev space of square-integrable functions with square-integrable gradients.  The finite element solution follows the standard Galerkin methodology \cite{zienkiewicz2005finite}. On each element $e$ we approximate the temperature field by:
\begin{equation}
T \approx T^h = \sum_{b\in\mathcal I_e} N_b\,T_b
\end{equation}
where $N_b$ are $C^0$ shape functions and $T_b$ are the nodal temperatures. The corresponding test function on element $e$ is taken as $v = N_a$, with $\mathcal I_e$ the set of local node indices. Then the element equations read:
\begin{equation}\label{eq:FE_WeakForm}
\sum_{b\in\mathcal I_e}
\left[
  \int_{\Omega^e} \mathbf{G}_a^\mathsf{T} \,\mathbf{k}^e\,\mathbf{G}_b\,\mathrm{d}\Omega
  \;+\;
  \int_{\Gamma_R^e} h\,N_a\,N_b\,\mathrm{d}\Gamma
\right] T_b
=
-\int_{\Omega^e} N_a\,f\,\mathrm{d}\Omega
\;+\;
\int_{\Gamma_N^e} N_a\,\bar{q}\,\mathrm{d}\Gamma
\;+\;
\int_{\Gamma_R^e} h\,T_{\infty}\,N_a\,\mathrm{d}\Gamma
\end{equation}
for all $a\in\mathcal I_e$, where $\mathbf{G}_a = \nabla N_a$, and $\mathbf k^e$ is the anisotropic conductivity tensor on element $e$, obtained from SPTC.

The finite-element system is assembled and solved using PETSc through the \texttt{DOLFINx} backends. After defining the UFL bilinear (LHS) and linear (RHS) forms of Equation ~\ref{eq:FE_WeakForm}, \texttt{DOLFINx} assembles the corresponding sparse linear system $A\mathbf{T}=\mathbf{b}$, applying essential (Dirichlet) boundary conditions by modifying the appropriate rows of $A$ and entries of $b$. Natural (Neumann) or mixed boundary conditions that would otherwise yield a singular system are regularized by pinning a single degree of freedom to remove the null space. In this work, the assembled system is then solved using PETSc via a direct sparse LU factorization with no Krylov iterations~\cite{KrylovSolverBook}. The resulting solution vector is returned as a \texttt{DOLFINx} Function and subsequently used to compute derived quantities such as $\nabla T^h$, the element heat flux $\mathbf{q}^e=-\sum_e\mathbf{k}^e\nabla T_b$, and the effective conductivity $K_\mathrm{eff}^\alpha$ obtained from face-integrated fluxes in the $\alpha$ thermal drive direction. All components of the solver were implemented through the Python interfaces of the FEniCS/DOLFINx and PETSc libraries.

\subsection{SPTC $\rightarrow$ FEM Nodal Conductivity (Voxelization)}\label{subsubsec:voxelization}
The \texttt{sptc2fem} module constructs the spatial conductivity field $\mathbf{K}(\mathbf{x})$ by mapping either computed or predicted SPTC values onto a hexahedral voxel grid. For $N_{\mathrm{at}}$ atoms with Cartesian positions $\{\mathbf{r}_a\}_{a=1}^{N_{\mathrm{at}}}$ and SPTC components $\zeta_a^{\beta}$, where $\beta\in\{x,y,z\}$ denotes the Cartesian components and $\beta=\mathrm{d}$ the isotropic value, we use $\zeta$ to denote either the true or predicted SPTC. The simulation supercell is defined by lattice vectors $\mathbf{k},\mathbf{l},\mathbf{m}\in\mathbb{R}^3$ forming the matrix $M=[\,\mathbf{k}\ \mathbf{l}\ \mathbf{m}\,]$ with Gram matrix $G = M^{\top}M$. Fractional coordinates, which satisfy periodic boundary conditions, are obtained as:
\begin{equation}
    \boldsymbol{\xi}_a = M^{-1}\mathbf{r}_a \in [0,1)^3
\end{equation}
with wrapping into $[0,1)^3$ applied component-wise as needed.

A regular voxel grid with resolution $N_{\mathrm{div}}=(n_x,n_y,n_z)$ is constructed in fractional space. Voxel centers (cell) and mesh vertices are defined by:
\begin{align}
\Xi^{\mathrm{cell}}_{ijk}
&= \Big(\tfrac{i+\tfrac{1}{2}}{n_x},\,\tfrac{j+\tfrac{1}{2}}{n_y},\,\tfrac{k+\tfrac{1}{2}}{n_z}\Big)
\\
\Xi^{\mathrm{vert}}_{IJK}
&= \Big(\tfrac{I}{n_x},\,\tfrac{J}{n_y},\,\tfrac{K}{n_z}\Big)
\end{align}
with indices $i=0,\dots,n_x-1$, $j=0,\dots,n_y-1$, $k=0,\dots,n_z-1$ for voxel centers, and $I=0,\dots,n_x$, $J=0,\dots,n_y$, $K=0,\dots,n_z$ for mesh vertices. Each voxel $(i,j,k)$ is bounded by the eight neighboring vertices $(I,J,K)\in\{i,i+1\}\times\{j,j+1\}\times\{k,k+1\}$, with periodic wrapping at the boundaries.

The SPTC values are interpolated to voxel centers using a Gaussian smoothing kernel of fractional width $h_{\mathrm{frac}}$:
\begin{align}
k^{\mathrm{cell},\beta}_{ijk}
&=
\frac{\displaystyle\sum_{a=1}^{N_{\mathrm{at}}}
       \mathcal{G}^{\mathrm{cell}}_{ijk;a}\,\zeta^\beta_a}
     {\displaystyle\sum_{a=1}^{N_{\mathrm{at}}}
       \mathcal{G}^{\mathrm{cell}}_{ijk;a}}
\\[3pt]
\mathcal{G}^{\mathrm{cell}}_{ijk;a}
&=
\exp\!\Big[
  -\frac{\big\|\Xi^{\mathrm{cell}}_{ijk}-\boldsymbol{\xi}_a\big\|_2^2}
        {2h_{\mathrm{frac}}^2}
\Big]
\\[3pt]
h_{\mathrm{frac}} &=
 s \times \min\!\left(\tfrac{1}{n_x},\tfrac{1}{n_y},\tfrac{1}{n_z}\right)
\end{align}
where $s=0.75$ controls the smoothing extent. Here the kernel is defined in fractional space; lattice-aware variants based on the Gram matrix $G$ are introduced later in the masking and void trimming analysis (see discussion in Section \textcolor{blue}{S2} \cite{SM}).

The conductivity at mesh vertices is obtained as the average over the surrounding voxel centers. For this, we denote $\mathcal{N}(I,J,K)$ as the set of voxel indices, such that:
\begin{equation}
\mathcal{N}(I,J,K)
=
\bigl\{(i,j,k)
  \mid i\in\{I-1,I\},\,
       j\in\{J-1,J\},\,
       k\in\{K-1,K\}
\bigr\}
\end{equation}
with all indices interpreted modulo $(n_x,n_y,n_z)$ to enforce periodic boundary conditions. Then:
\begin{equation}\label{eq:voxelization}
k^{\mathrm{vert},\beta}_{IJK}
=
\frac{1}{|\mathcal{N}(I,J,K)|}
\sum_{(i,j,k)\in\mathcal{N}(I,J,K)}
k^{\mathrm{cell},\beta}_{ijk}
\end{equation}
For directional SPTC components we define, at each mesh vertex $(I,J,K)$, a diagonal nodal conductivity tensor:
\begin{equation}\label{eq:K_node}
\mathbf{K}^{\mathrm{node}}_{IJK}
=
\mathrm{diag}\!\big(k^{\mathrm{vert},x}_{IJK},\,k^{\mathrm{vert},y}_{IJK},\,k^{\mathrm{vert}, z}_{IJK}\big)
\end{equation}
For purely isotropic SPTC, we set $k^{\mathrm{vert},x}_{IJK}=k^{\mathrm{vert},y}_{IJK}=k^{\mathrm{vert},z}_{IJK}=k^{\mathrm{vert},\mathrm{d}}_{IJK}$ and obtain $\mathbf{K}^{\mathrm{node}}_{IJK}=k^{\mathrm{vert},\mathrm{d}}_{IJK}\times\mathbf{I}$.

If $\mathcal{I}_e$ denotes the set of vertices (nodes) of element $e$ with index $n$, we construct the anisotropic conductivity tensor on each element by averaging the nodal tensors:
\begin{equation}\label{eq:Ke_from_nodes}
\mathbf{K}^e
= \frac{1}{|\mathcal{I}_e|}
  \sum_{n\in\mathcal{I}_e}  \mathbf{K}^{\mathrm{node}}_\mathrm{n}
\end{equation}
In this way, the SPTC-derived nodal conductivities $ \mathbf{K}^{\mathrm{node}}_\mathrm{n}$ define the element conductivity tensors $\mathbf{K}^e$ that enter the discretized constitutive relation $\mathbf{q}^e = -\mathbf{K}^e \nabla T^h$ in each element. Visualization of the FE models was done using Paraview \cite{ParaView}.

\subsection{Effective Conductivity and Conductivity Scaling}\label{subsubsec:Keff}

Following voxelization, the DOLFINx-based FEM solver estimates homogenized (or effective) conductivity components $K_{\mathrm{eff}}^\alpha$ from steady-state heat solves along a given drive (thermal loading) direction,  \(\hat{\boldsymbol{\alpha}}\in{\hat{\mathbf{x}},\hat{\mathbf{y}},\hat{\mathbf{z}}}\). For a given $\alpha$ with the RVE spanning a length $D_\alpha$ along the drive direction, \texttt{SCACS} solves a two-face (\emph{left--right}) boundary value problem by prescribing temperatures on the left ($L$) and right ($R$) planes, with $f(\mathbf{x})=0$. The effective conductivity in direction $\alpha$ is then computed from the face-integrated fluxes as:
\begin{equation}\label{eq:K_eff_flux}
K_\mathrm{eff}^\alpha
= \frac{\overline{|Q|}}{\overline{A}}\;\frac{D_\alpha}{|\Delta T|}
\end{equation}
where $\Delta T = T_R - T_L$ is the applied temperature difference and:
\begin{equation}
\overline{|Q|}
= \tfrac{1}{2}\bigl(|Q_L|+|Q_R|\bigr);\qquad
\overline{A}
= \tfrac{1}{2}\bigl(A_L + A_R\bigr)
\end{equation}
denote the average magnitude of the normal heat flux and the average face area on the two driven boundaries, respectively. Here \( Q_\gamma = \int_{\gamma} q_n\,\mathrm{d}\gamma\) is the total normal heat flux through face $\gamma\in\{L,R\}$, with $q_n = \mathbf{n}^\mathsf{T}(\mathbf{K}\nabla T)$ the normal component of the local flux and $\mathbf{n}$ the outward unit normal.

When a target (experimental) bulk conductivity $k_\mathrm{target}$ is prescribed, \texttt{SCACS} first performs a Dirichlet calibration solve using the SPTC-derived nodal conductivity field, $\mathbf{K}^{\mathrm{node}}_\mathrm{n}$. From this calibration solution we compute a volume-averaged effective conductivity in direction $\alpha$:
\begin{equation}\label{eq:K_eff_vol}
K_\mathrm{eff}^{\mathrm{vol},\alpha}
= \frac{1}{|\Omega|}
  \int_{\Omega}
    (K^{\mathrm{node}}_{\mathrm{n}}\nabla T)\!\cdot\!\hat{\boldsymbol{\alpha}}
  \,\mathrm{d}\Omega
  \;\times\;
  \frac{D_\alpha}{\Delta T}.
\end{equation}
This expression is the standard continuum volume identity for the effective conductivity in direction $\hat{\boldsymbol{\alpha}}$, up to the normalization factor $D_\alpha / \Delta T$ and the choice of drive direction used in the calibration solve (here with $\Delta T=1$ for convenience).

We then apply a uniform rescaling to obtain a calibrated conductivity field $\mathbf{K}_\mathrm{s}(\mathbf{x})$:
\begin{equation}\label{eq:K_scaled}
  \mathbf{K}_\mathrm{s}(\mathbf{x})
  = \frac{k_\mathrm{target}}{K_\mathrm{eff}^{\mathrm{vol},\alpha}}\,
    \mathbf{K}_{\mathrm{n}}(\mathbf{x}).
\end{equation}
Because the governing equation is linear in $\mathbf{K}(\mathbf{x})$, this global scaling factor rescales both the volume-based and flux-based effective conductivities in direction $\alpha$ so that $K_\mathrm{eff}^\alpha \approx k_\mathrm{target}$, up to discretization error. All subsequent finite-element solves and homogenization steps use $\mathbf{K}_\mathrm{s}(\mathbf{x})$,  while preserving the spatial anisotropy encoded by the original SPTC-derived field.

\subsection{Adaptive Mesh Refinement (AMR)}\label{subsubsec:amr-sptc}

Adaptive mesh refinement (AMR) in \texttt{SCACS} is driven by a residual-based error indicator that quantifies local imbalance in the discrete heat-flux field. For each hexahedral cell $\mathcal{K}$ with volumetric source $f(\mathbf{x})$, the indicator is defined as:
\begin{equation}
\eta_{\mathcal{K}} = h_\mathcal{K}\,\big|\overline{f}_\mathcal{K} + (\nabla\!\cdot\!\mathbf{q})_\mathcal{K}\big|
\end{equation}
where $\overline{f}_\mathcal{K}$ is the cell-averaged source term and $h_\mathcal{K}=\tfrac{1}{3}(\Delta x+\Delta y+\Delta z)$ is the characteristic cell size. For anisotropic materials, the effective size is scaled by the directional conductivities as:
\begin{equation}
h_\mathcal{K}^* = \frac{1}{3}\!\left(
\frac{\Delta x}{\sqrt{K^x_\mathcal{K}/K_\mathrm{ref}}} +
\frac{\Delta y}{\sqrt{K^y_\mathcal{K}/K_\mathrm{ref}}} +
\frac{\Delta z}{\sqrt{K^z_\mathcal{K}/K_\mathrm{ref}}}
\right)
\end{equation}
where $K^\alpha_\mathcal{K}$ is the $\mathcal{K}$-averaged conductivity in $\alpha$-direction  and \(K_\mathrm{ref}=\tfrac{1}{3}(K^x_\mathcal{K}+K^y_\mathcal{K}+K^z_\mathcal{K})\). In the isotropic limit, $h_\mathcal{K}^* \!\rightarrow\! h_\mathcal{K}$. The divergence of the $\alpha$-component of the flux is approximated by centered face differences as:
\begin{equation}
\partial_\alpha q^\alpha \approx 
\frac{\overline{q}^{(+\alpha)}-\overline{q}^{(-\alpha)}}{\Delta \alpha}
\end{equation}

\texttt{SCACS} implements the regenerative h-refinement where mesh refinement is achieved by globally redistributing the grid planes to concentrate resolution in regions where $\eta_\mathcal{K}$ is large. Since an optimal mesh is one where the distribution of element residual error is equal for all elements \cite{zienkiewicz2005finite}, we introduce the \emph{Equal-Mass Controller} (EMC) that ensures redistribution of grid edges based on the local element residual. The EMC retains the mathematical principles of AMR while remaining consistent with the voxelized, lattice-aligned SPTC fields.

\subsubsection{The Equal-Mass Controller (EMC):}\label{Par:EMC}
The EMC realizes this structured-grid AMR strategy by adaptively repositioning the one-dimensional edge arrays \((e^x, e^y, e^z)\) to concentrate resolution in regions of high residual error. All refinements operate on the same cell-wise indicator set \(\{\eta_\mathcal{K}\}\). The EMC redistributes grid edges such that the cumulative per-axis ``cell mass'' defined as:
\begin{equation}
m_\mathcal{K} = \eta_\mathcal{K} V_\mathcal{K}
\end{equation}
is equi-distributed along each coordinate direction. Here, \(V_\mathcal{K}\) denotes the Cartesian volume of the \(\mathcal{K}^{\mathrm{th}}\) cell. 
A D{\"o}rfler-type bulk marking strategy is employed to bias refinement toward a D{\"o}rfler region of interest (ROI) \cite{Dorfler1996}.  Both the equal-mass redistribution and ROI bias respect lattice geometry, boundary alignment, and refinement budgets, and are constrained by prescribed minimum/maximum element sizes and slope-ratio limits, all detailed below.

For grid budgeting, the current grid resolution is denoted \(N_0 = (n_x, n_y, n_z)\) with a total of \(n_x \times n_y \times n_z\) cells.  Given a growth cap \(g_{\mathrm{cells}}>0\) on the total number of cells, the per-axis scale factor is \(g_{\mathrm{cells}}^{1/3}\), yielding the (new) target resolution:
\begin{equation}
N' = (n'_x, n'_y, n'_z) \approx g_{\mathrm{cells}}^{1/3}\,N_0
\end{equation}

EMC then ``flattens'' the three-dimensional mesh onto each coordinate axis to determine where new grid planes should be inserted. For a chosen axis \(a \in \{x, y, z\}\) spanning \([a_{\min}, a_{\max}]\), each cell \(\mathcal{K}\) is projected onto the interval
\begin{equation}
[a^{\mathrm{cell}}_{\min,\mathcal{K}}, a^{\mathrm{cell}}_{\max,\mathcal{K}}];
\qquad
\ell^a_\mathcal{K} = a^{\mathrm{cell}}_{\max,\mathcal{K}} - a^{\mathrm{cell}}_{\min,\mathcal{K}}
\end{equation}
which identifies the cell’s location between the left and right boundaries when viewed along that axis.

All such cell faces are collected together with the global endpoints \(a_{\min}\) and \(a_{\max}\) to form a unique, ordered set of potential projection planes:
\begin{equation}
\Pi = \{a_{\min}, a_{\max}\} 
\cup \{a^{\mathrm{cell}}_{\min,\mathcal{K}}, a^{\mathrm{cell}}_{\max,\mathcal{K}}\}_\mathcal{K}
\end{equation}
The ordered array \(\Pi = \{\Pi_0, \Pi_1, \dots, \Pi_S\}\) defines all distinct planes perpendicular to the axis. The intervals between successive planes, \([\Pi_j, \Pi_{j+1}]\) of width \(\Delta a_j = \Pi_{j+1} - \Pi_j\), constitute the elementary one-dimensional segments over which the controller integrates the projected ``mass'' \((\eta_\mathcal{K} V_\mathcal{K})\) to determine how resolution should be redistributed along that direction.

Each cell mass is uniformly distributed along its axis span, yielding a piecewise-constant projected density
\begin{equation}
\rho_j = 
\sum_\mathcal{c} 
\frac{m_\mathcal{K}}{\ell^a_\mathcal{K}}
\end{equation}
where the summation condition $\mathcal{c}$ implies ${\mathcal{K} : [\Pi_j,\Pi_{j+1}] \subset 
[a^{\mathrm{cell}}_{\min,\mathcal{K}}, a^{\mathrm{cell}}_{\max,\mathcal{K}}]}$. The mass of each segment is then  \(M_{\mathrm{s}}(j) = \rho_j \Delta a_j\), and the total projected mass is \(M_{\mathrm{tot}} = \sum_j M_{\mathrm{s}}(j)\).

From these, a cumulative distribution function (CDF) is constructed, initialized as \(M(\Pi_0)=0\):
\begin{equation}
M(\Pi_{j+1}) = M(\Pi_j) + M_{\mathrm{s}}(j)
\end{equation}
New edge coordinates \(\tilde{e}^a_k\) are positioned such that each cell encloses an equal fraction of the total mass:
\begin{equation}
\tilde{e}^\alpha_k = 
M^{-1}\!\left(\frac{k}{n'_\alpha}\,M_{\mathrm{tot}}\right),
\qquad k = 0, \dots, n'_\alpha
\end{equation}

To maintain smooth gradation, the following slope-ratio limiting step enforces constraints on adjacent cell widths while preserving the total domain length. The boundary planes are fixed such that \(\tilde{e}^a_0 = a_{\min}\) and  \(\tilde{e}^a_{n'_a} = a_{\max}\). Defining interval widths \(h_i = \tilde{e}^a_{i+1} - \tilde{e}^a_i\), we impose upper and lower bounds (\(h_{\min}, h_{\max}\)) and a slope constraint ratio \(r \ge 1\), so that:
\begin{gather}
h_{\min} \le h_i \le h_{\max} \label{eq:slopeLimiting1}\\ 
\frac{1}{r} \le \frac{h_{i+1}}{h_i} \le r \label{eq:slopeLimiting2}\\
\sum_i h_i = a_{\max} - a_{\min} \label{eq:slopeLimiting3}
\end{gather}
The complete equal-mass refinement is performed independently for each axis \(a \in \{x, y, z\}\), yielding refined edge arrays \((\tilde{e}^x, \tilde{e}^y, \tilde{e}^z)\).

\subsubsection{D{\"o}rfler-Type ROI Bias:}\label{par:DorflerROI}
To concentrate refinement within regions of dominant error, we adopt a D{\"o}rfler-type bulk-marking strategy~\cite{Dorfler1996} to define a one-dimensional region of interest (ROI) along each coordinate axis. 
For each cell \(\mathcal{K}\), a local score is computed as: 
\begin{equation}
s_\mathcal{K} = \eta_\mathcal{K}^2 V_\mathcal{K}
\end{equation}
and a minimal marked set \(\mathcal{M}\) is selected such that \cite{Dorfler1996}:
\begin{equation}
\sum_{\mathcal{K}\in\mathcal{M}} s_\mathcal{K} 
\ge 
d \sum_{\mathcal{K}} s_\mathcal{K}
\end{equation}
ensuring that the selected elements (cells) account for a fixed fraction \(d \in (0,1)\) of the total residual energy. Projecting the marked cells \(\mathcal{M}\) onto the \(\alpha\)-axis yields the spatial extent of the ROI, bounded by \([\ell^*_\alpha, r^*_\alpha]\).  where $\ell^*_\alpha$ and $r^*_\alpha$ are the raw (unpadded) endpoints of the high-error region along axis. $\ell^*_\alpha$ is the smallest coordinate on axis $\alpha$ reached by any marked cell (the leftmost extent of $\mathcal{M}$ when projecting its cells onto $\alpha$), and $r^*_\alpha$ is the largest (rightmost) counterpart. 

Defining the domain length as \(L_\alpha = \alpha_{\max} - \alpha_{\min}\), the ROI is padded symmetrically by a fractional width \(p \in (0,1)\), so that:
\begin{equation}
\mathrm{ROI}^\alpha = [\ell_\alpha, r_\alpha] 
= [\max(\alpha_{\min},\, \ell^*_\alpha - pL_\alpha),\, \min(\alpha_{\max},\, r^*_\alpha + pL_\alpha)]
\end{equation}

The axis is then divided into three contiguous segments: a left region outside the ROI ($L_\mathrm{L}$), the middle ROI ($L_\mathrm{M}$), and a right region outside the ROI ($L_\mathrm{R}$):
\begin{align}
L_\mathrm{L} &= \ell_\alpha - \alpha_{\min}\\
L_\mathrm{M} &= r_\alpha - \ell_\alpha\\
L_\mathrm{R} &= \alpha_{\max} - r_\alpha\\
L_{\mathrm{T}} &= L_\mathrm{L} + L_\mathrm{M} + L_\mathrm{R}
\end{align}
The segment length fractions are then:
\begin{equation}
f_{\mathrm{j}} = \frac{L_{\mathrm{j}}}{L_{\mathrm{T}}}; \quad
\mathrm{j} \in \{\mathrm{L}, \mathrm{M}, \mathrm{R}\}
\end{equation}
To allocate a greater share of refinement to the ROI, the middle fraction is boosted by a factor \((1+\omega)\), where \(\omega > 0\):
\begin{equation}
f_\mathrm{M}^\mathrm{eff} = \min\{1, (1+\omega)f_\mathrm{M}\}
\end{equation}
This scaling ensures that the ROI receives a larger portion of the available cell budget, concentrating resolution where it is most needed.  The outer fractions are then renormalized to maintain the total fraction sum:
\begin{equation}
f_\mathrm{L}^\mathrm{eff} : f_\mathrm{R}^\mathrm{eff} = f_\mathrm{L} : f_\mathrm{R}\quad \mathrm{s.t.}\quad \sum_j f^\mathrm{eff}_j = 1
\end{equation}
The number of target cells per region along axis \(\alpha\) is then computed as:
\begin{align}
N_\mathrm{L} &= f_\mathrm{L}^\mathrm{eff} n'_\alpha\\
N_\mathrm{M} &= \max\{1,\,f_\mathrm{M}^\mathrm{eff} n'_\alpha\}\\
N_\mathrm{R} &= n'_\alpha - N_\mathrm{L} - N_\mathrm{M}
\end{align}
where \(n'_\alpha\) is the total target cell count along that axis. If either outer region is non-empty but under-resolved, a minimum of \(N_\mathrm{min} \ge 1\) cells is enforced, and \(N_\mathrm{M}\) is adjusted accordingly.

Each sub-interval---the left \([\alpha_{\min}, \ell_\alpha]\), middle (ROI) \([\ell_\alpha, r_\alpha]\), and right \([r_\alpha, \alpha_{\max}]\)---is then refined independently via EMC (Section~\ref{Par:EMC}). The resulting edge arrays are concatenated sequentially, omitting duplicate interfaces at \(\ell_\alpha\) and \(r_\alpha\), to form a single monotonic edge list for the full axis. Finally, the combined edges undergo slope limiting and normalization (Equations~\ref{eq:slopeLimiting1}--\ref{eq:slopeLimiting3}) to ensure smooth gradation of cell widths and strict preservation of the total domain length.

\subsubsection{AMR Stopping Criterion:}\label{par:AMRCriterion}
The adaptive refinement loop terminates once the flux effective conductivity, $\tau_\mathrm{K}$ (Equation \ref{eq:K_eff_flux}) and the residual indicator ($\tau_\eta$) are both stable between consecutive iterations $k-1$ and $k$ as:
\begin{align}\label{eq:AMRcirteria}
\tau^*_\mathrm{K} &\ge \frac{|K_{\mathrm{eff}}^{(k)}-K_{\mathrm{eff}}^{(k-1)}| }
     {\max\{|K_{\mathrm{eff}}^{(k)}|,\varepsilon\}} \equiv \tau_\mathrm{K} \\[3pt]
\tau^*_\eta &\ge \frac{|\|\eta^{(k)}\|_2-\|\eta^{(k-1)}\|_2|}
     {\max\{\|\eta^{(k)}\|_2,\varepsilon\}} \equiv \tau_\eta 
\end{align}
with $\varepsilon = 10^{-30}$, and typical tolerances set as: $\tau^*_\mathrm{K} = 5\times10^{-3}$ and $\tau^*_\eta = 0.1$.

\bibliography{references}

@PREAMBLE{
 "\providecommand{\noopsort}[1]{}" 
 # "\providecommand{\singleletter}[1]{#1}%" 
}

@misc{SM,
    note = {See Supplemental Material at [URL will be inserted by publisher] for additional details on the SPTC graph network, the masking procedure used to generate FE models with voids, and the atomic and FE model datasets discussed in the paper.}
}

@article{Zenodo,
    author = {Ugwumadu, Chinonso and Drabold,  David A and Tutchton, Roxanne M},
    title = {Seamlessly joining length scales: From atomistic thermal graphs to anisotropic continuum conductivity [Data set]},
    journal = {Zenodo},
    year = {2026},
    doi={10.5281/zenodo.18474574}
}

@article{Takeuchi2020PolySi,
doi = {10.35848/1347-4065/ab9624},
year = {2020},
month = {jun},
publisher = {IOP Publishing},
volume = {59},
number = {7},
pages = {075501},
author = {Takeuchi, Haruki and Yokogawa, Ryo and Takahashi, Kazuya and Komori, Katsuhiko and Morimoto, Tamotsu and Yamashita, Yuichiro and Sawamoto, Naomi and Ogura, Atsushi},
title = {Thermal conductivity characteristics in polycrystalline silicon with different average sizes of grain and nanostructures in the grains by UV Raman spectroscopy},
journal = {Japanese Journal of Applied Physics}
}

@article{DavisNanopillars2014,
  title = {Nanophononic Metamaterial: Thermal Conductivity Reduction by Local Resonance},
  author = {Davis, Bruce L. and Hussein, Mahmoud I.},
  journal = {Phys. Rev. Lett.},
  volume = {112},
  issue = {5},
  pages = {055505},
  numpages = {5},
  year = {2014},
  month = {Feb},
  publisher = {American Physical Society},
  doi = {10.1103/PhysRevLett.112.055505}
}

@article{Boukai2008nanowires,
  title={Silicon nanowires as efficient thermoelectric materials},
  author={Boukai, Akram I and Bunimovich, Yuri and Tahir-Kheli, Jamil and Yu, Jen-Kan and Goddard Iii, William A and Heath, James R},
  journal={nature},
  volume={451},
  number={7175},
  pages={168--171},
  year={2008},
  doi={10.1038/nature06458},
  publisher={Nature Publishing Group UK London}
}

@article{hochbaum2008nanowires,
  title={Enhanced thermoelectric performance of rough silicon nanowires},
  author={Hochbaum, Allon I and Chen, Renkun and Delgado, Raul Diaz and Liang, Wenjie and Garnett, Erik C and Najarian, Mark and Majumdar, Arun and Yang, Peidong},
  journal={Nature},
  volume={451},
  number={7175},
  pages={163--167},
  year={2008},
  doi={10.1038/nature06381},
  publisher={Nature Publishing Group UK London}
}

@article{Kang2017GKF,
  title = {First-principles Green-Kubo method for thermal conductivity calculations},
  author = {Kang, Jun and Wang, Lin-Wang},
  journal = {Phys. Rev. B},
  volume = {96},
  issue = {2},
  pages = {020302},
  numpages = {5},
  year = {2017},
  month = {Jul},
  publisher = {American Physical Society},
  doi = {10.1103/PhysRevB.96.020302}
}

@article{JJ2021,
author = {Zeng, Yi and Avritte, Jalaan T. and Dong, Jianjun},
title = {Generalized Langevin Equation Theory of Thermal Conduction across Material Interfaces},
journal = {physica status solidi (b)},
volume = {258},
number = {9},
pages = {2000454},
doi = {10.1002/pssb.202000454},
year = {2021}
}

@article{Schelling2002Comparison,
  title = {Comparison of atomic-level simulation methods for computing thermal conductivity},
  author = {Schelling, Patrick K. and Phillpot, Simon R. and Keblinski, Pawel},
  journal = {Phys. Rev. B},
  volume = {65},
  issue = {14},
  pages = {144306},
  numpages = {12},
  year = {2002},
  month = {Apr},
  publisher = {American Physical Society},
  doi = {10.1103/PhysRevB.65.144306}
}

@article{Maginn2018EMD, 
title={Best Practices for Computing Transport Properties 1. Self-Diffusivity and Viscosity from Equilibrium Molecular Dynamics}, 
volume={1}, 
DOI={10.33011/livecoms.1.1.6324}, 
number={1}, 
journal={Living Journal of Computational Molecular Science}, 
author={Maginn, Edward J. and Messerly, Richard A. and Carlson, Daniel J. and Roe, Daniel R. and Elliot, J. Richard}, year={2018},
month={Dec.}, 
pages={6324} 
}

@book{McQuarrie1997,
  title={Statistical mechanics University Science Books},
  author={McQuarrie, DA},
  year={1997},
  page={641},
  publisher={New York: Harper and Row}
}

@article{MuellerPlathe1997NEMD,
    author = {Müller-Plathe, Florian},
    title = {A simple nonequilibrium molecular dynamics method for calculating the thermal conductivity},
    journal = {The Journal of Chemical Physics},
    volume = {106},
    number = {14},
    pages = {6082-6085},
    year = {1997},
    month = {04},
    issn = {0021-9606},
    doi = {10.1063/1.473271}
}

@article{Zhou2015NEMD,
  title = {Quantitatively analyzing phonon spectral contribution of thermal conductivity based on nonequilibrium molecular dynamics simulations. II. From time Fourier transform},
  author = {Zhou, Yanguang and Hu, Ming},
  journal = {Phys. Rev. B},
  volume = {92},
  issue = {19},
  pages = {195205},
  numpages = {10},
  year = {2015},
  month = {Nov},
  publisher = {American Physical Society},
  doi = {10.1103/PhysRevB.92.195205}
}

@article{Yang2013MFP,
  title = {Mean free path spectra as a tool to understand thermal conductivity in bulk and nanostructures},
  author = {Yang, Fan and Dames, Chris},
  journal = {Phys. Rev. B},
  volume = {87},
  issue = {3},
  pages = {035437},
  numpages = {12},
  year = {2013},
  month = {Jan},
  publisher = {American Physical Society},
  doi = {10.1103/PhysRevB.87.035437}
}

@article{Koh2007,
  title = {Frequency dependence of the thermal conductivity of semiconductor alloys},
  author = {Koh, Yee Kan and Cahill, David G.},
  journal = {Phys. Rev. B},
  volume = {76},
  issue = {7},
  pages = {075207},
  numpages = {5},
  year = {2007},
  month = {Aug},
  publisher = {American Physical Society},
  doi = {10.1103/PhysRevB.76.075207}
}

@article{Larkin2014MFP,
  title = {Thermal conductivity accumulation in amorphous silica and amorphous silicon},
  author = {Larkin, Jason M. and McGaughey, Alan J. H.},
  journal = {Phys. Rev. B},
  volume = {89},
  issue = {14},
  pages = {144303},
  numpages = {12},
  year = {2014},
  month = {Apr},
  publisher = {American Physical Society},
  doi = {10.1103/PhysRevB.89.144303}
}

@article{Broido2007BOLTZ,
    author = {Broido, D. A. and Malorny, M. and Birner, G. and Mingo, Natalio and Stewart, D. A.},
    title = {Intrinsic lattice thermal conductivity of semiconductors from first principles},
    journal = {Applied Physics Letters},
    volume = {91},
    number = {23},
    pages = {231922},
    year = {2007},
    month = {12},
    issn = {0003-6951},
    doi = {10.1063/1.2822891}
}

@article{Lindsay2016BOLTZ,
author = {Lucas Lindsay},
title = {First Principles Peierls-Boltzmann Phonon Thermal Transport: A Topical Review},
journal = {Nanoscale and Microscale Thermophysical Engineering},
volume = {20},
number = {2},
pages = {67--84},
year = {2016},
publisher = {Taylor \& Francis},
doi = {10.1080/15567265.2016.1218576}
}

@article{Nepal2025GST,
title = {Electronic and thermal properties of the phase-change memory material, Ge2Sb2Te5, and results from spatially resolved transport calculations},
journal = {Solid State Sciences},
volume = {173},
pages = {108182},
year = {2026},
issn = {1293-2558},
doi = {10.1016/j.solidstatesciences.2025.108182},
author = {K. Nepal and A. Gautam and R. Hussein and K. Konstantinou and S.R. Elliott and C. Ugwumadu and D.A. Drabold},
}

@article{Ugwumadu2024Cfoam,
title = {Atomistic-to-continuum modeling of carbon foam: A new approach to finite element simulation},
journal = {Carbon},
volume = {229},
pages = {119506},
year = {2024},
issn = {0008-6223},
doi = {10.1016/j.carbon.2024.119506},
author = {C. Ugwumadu and W. Downs and C. O’Brien and R. Thapa and R. Olson and B. Wisner and M. Ali and J. Trembly and Y. Al-Majali and D.A. Drabold}
}

@article{Hill1972O,
  title={On constitutive macro-variables for heterogeneous solids at finite strain},
  author={Rodney Hill},
  journal={Proceedings of the Royal Society of London. A. Mathematical and Physical Sciences},
  year={1972},
  volume={326},
  pages={131 - 147},
  doi={10.1098/rspa.1972.0001}
}

@book{mandel1972,
  title={Plasticite classique et viscoplasticite: },
  author={Mandel, J.},
  isbn={9780387811970},
  series={CISM International Centre for Mechanical Sciences},
  year={1972},
  publisher={Springer}
}

@article{Roberto2017,
author = {Ramos, Gustavo Roberto and dos Santos, Tiago and Rossi, Rodrigo},
title = {An extension of the Hill–Mandel principle for transient heat conduction in heterogeneous media with heat generation incorporating finite RVE thermal inertia effects},
journal = {International Journal for Numerical Methods in Engineering},
volume = {111},
number = {6},
pages = {553-580},
doi = {10.1002/nme.5471},
year = {2017}
}

@article{Dorfler1996,
author = {D\"{o}rfler, Willy},
title = {A Convergent Adaptive Algorithm for Poisson’s Equation},
journal = {SIAM Journal on Numerical Analysis},
volume = {33},
number = {3},
pages = {1106-1124},
year = {1996},
doi = {10.1137/0733054}
}

@article{Bartok2010GAP,
  title = {Gaussian Approximation Potentials: The Accuracy of Quantum Mechanics, without the Electrons},
  author = {Bart\'ok, Albert P. and Payne, Mike C. and Kondor, Risi and Cs\'anyi, G\'abor},
  journal      = {Physical Review Letters},
  volume = {104},
  issue = {13},
  pages = {136403},
  numpages = {4},
  year = {2010},
  month = {Apr},
  publisher = {American Physical Society},
  doi = {10.1103/PhysRevLett.104.136403}
}

@article{loshchilov2019,
    author = {Ilya Loshchilov and Frank Hutter},
    title = {Decoupled Weight Decay Regularization},
    journal = {arXiv},
    year = {2019},
    doi = {arXiv:1711.05101v3}
}

@article{Gonzalez1985FPS,
title = {Clustering to minimize the maximum intercluster distance},
journal = {Theoretical Computer Science},
volume = {38},
pages = {293-306},
year = {1985},
issn = {0304-3975},
doi = {10.1016/0304-3975(85)90224-5},
author = {Teofilo F. Gonzalez},
}

@inproceedings{perez2018film,
  title={Film: Visual reasoning with a general conditioning layer},
  author={Perez, Ethan and Strub, Florian and De Vries, Harm and Dumoulin, Vincent and Courville, Aaron},
  booktitle={Proceedings of the AAAI conference on artificial intelligence},
  volume={32},
  number={1},
  year={2018}
}

@article{Gabriel2024,
author = {Barrenechea, Gabriel R. and John, Volker and Knobloch, Petr},
title = {Finite Element Methods Respecting the Discrete Maximum Principle for Convection-Diffusion Equations},
journal = {SIAM Review},
volume = {66},
number = {1},
pages = {3-88},
year = {2024},
doi = {10.1137/22M1488934}
}

@article{Uma2001,
  title={Temperature-dependent thermal conductivity of undoped polycrystalline silicon layers},
  author={Uma, S and McConnell, AD and Asheghi, M and Kurabayashi, K and Goodson, KE},
  journal={International Journal of Thermophysics},
  volume={22},
  number={2},
  pages={605--616},
  year={2001},
  publisher={Springer},
  doi={10.1023/A:1010791302387}
}

@article{Abel2007,
  title={Thermal metrology of silicon microstructures using Raman spectroscopy},
  author={Abel, Mark R and Wright, Tanya L and King, William P and Graham, Samuel},
  journal={IEEE Transactions on components and packaging technologies},
  volume={30},
  number={2},
  pages={200--208},
  year={2007},
  publisher={IEEE},
  doi={10.1109/TCAPT.2007.897993}
}

@article{Gautam2025,
author = {Gautam, Aashish and Lee, Yoon Gyu and Ugwumadu, Chinonso and Nepal, Kishor and Nakhmanson, Serge and Drabold, David A.},
title = {Site-Projected Thermal Conductivity: Application to Defects, Interfaces, and Homogeneously Disordered Materials},
journal = {physica status solidi (RRL) – Rapid Research Letters},
volume = {19},
number = {2},
pages = {2400306},
doi = {10.1002/pssr.202400306},
year = {2025}
}

@article{Ugwumadu2025,
author = {Ugwumadu, Chinonso and Gautam, Aashish and Lee, Yoon Gyu and Drabold, David A.},
title = {Mapping Thermal Conductivity at the Atomic Scale: A Step toward the Thermal Design of Materials},
journal = {physica status solidi (b)},
pages = {2500316},
year ={2025},
doi = {10.1002/pssb.202500316},
}

@article{UgwumaduPSSB2025,
author = {Ugwumadu, Chinonso and Drabold, David A. and Tutchton, Roxanne M.},
title = {Effects of Galactic Irradiation on Thermal and Electronic Transport in Tungsten},
journal = {physica status solidi (b)},
volume = {262},
number = {11},
pages = {2500109},
doi = {10.1002/pssb.202500109},
year = {2025}
}

@article{Powers,
    author = {Powers, Joseph M. },
    title = {On the Necessity of Positive Semi-Definite Conductivity and Onsager Reciprocity in Modeling Heat Conduction in Anisotropic Media },
    journal = {Journal of Heat Transfer},
    volume = {126},
    number = {5},
    pages = {670-675},
    year = {2004},
    month = {11},
    issn = {0022-1481},
    doi = {10.1115/1.1798913}
}

@article{rosset2025,
  title={Signatures of paracrystallinity in amorphous silicon from machine-learning-driven molecular dynamics},
  author={Rosset, Louise AM and Drabold, David A and Deringer, Volker L},
  journal={Nature Communications},
  volume={16},
  number={1},
  pages={2360},
  year={2025},
  doi={10.1038/s41467-025-57406-4},
  publisher={Nature Publishing Group UK London}
}

@article{SOAP,
  title = {On representing chemical environments},
  author = {Bart\'ok, Albert P. and Kondor, Risi and Cs\'anyi, G\'abor},
  journal = {Phys. Rev. B},
  volume = {87},
  issue = {18},
  pages = {184115},
  numpages = {16},
  year = {2013},
  month = {May},
  publisher = {American Physical Society},
  doi = {10.1103/PhysRevB.87.184115},
}

@InCollection{ParaView,
  author    = {James Ahrens and Berk Geveci and Charles Law},
  booktitle = {Visualization Handbook},
  publisher = {Elesvier},
  title     = {{ParaView}: An End-User Tool for Large Data Visualization},
  year      = {2005},
  note      = {{ISBN}~978-0123875822},
}

@article{Kubo_TC,
  title={Statistical-Mechanical Theory of Irreversible Processes. II. Response to Thermal Disturbance},
  author={Ryogo Kubo and Mario Yokota and Sadao Nakajima},
  journal={Journal of the Physical Society of Japan},
  volume={12},
  number={11},
  pages={1203-1211},
  year={1957},
  doi={10.1143/JPSJ.12.1203}
}

@article{Green_TC,
author = {Green,Melville S. },
title = {Markoff Random Processes and the Statistical Mechanics of Time‐Dependent Phenomena. II. Irreversible Processes in Fluids},
journal = {The Journal of Chemical Physics},
volume = {22},
number = {3},
pages = {398-413},
year = {1954},
doi = {10.1063/1.1740082}
}

@article{AF3,
  title = {Thermal conductivity of disordered harmonic solids},
  author = {Allen, Philip B. and Feldman, Joseph L.},
  journal = {Phys. Rev. B},
  volume = {48},
  issue = {17},
  pages = {12581--12588},
  numpages = {0},
  year = {1993},
  month = {Nov},
  publisher = {American Physical Society},
  doi = {10.1103/PhysRevB.48.12581}
}

@article{AF1,
  title = {Thermal Conductivity of Glasses: Theory and Application to Amorphous Si},
  author = {Allen, Philip B. and Feldman, Joseph L.},
  journal = {Phys. Rev. Lett.},
  volume = {62},
  issue = {6},
  pages = {645--648},
  numpages = {0},
  year = {1989},
  month = {Feb},
  publisher = {American Physical Society},
  doi = {10.1103/PhysRevLett.62.645}
}

@article{Hardy,
  title = {Energy-Flux Operator for a Lattice},
  author = {Hardy, Robert J.},
  journal = {Phys. Rev.},
  volume = {132},
  issue = {1},
  pages = {168--177},
  numpages = {0},
  year = {1963},
  month = {Oct},
  publisher = {American Physical Society},
  doi = {10.1103/PhysRev.132.168}
}

@article{lammps,
title = {{LAMMPS - a flexible simulation tool for particle-based materials modeling at the atomic, meso, and continuum scales}},
journal = {Computer Physics Communications},
volume = {271},
pages = {108171},
year = {2022},
issn = {0010-4655},
doi = {10.1016/j.cpc.2021.108171},
author = {{Aidan P. Thompson and H. Metin Aktulga and Richard Berger and Dan S. Bolintineanu and W. Michael Brown and Paul S. Crozier and Pieter J. {in `t Veld} and Axel Kohlmeyer and Stan G. Moore and Trung Dac Nguyen and Ray Shan and Mark J. Stevens and Julien Tranchida and Christian Trott and Steven J. Plimpton}},
}

@article{Uguwmadu2024NPC,
author = {Ugwumadu, C. and Thapa, R. and Nepal, K. and Gautam, A. and Al-Majali, Y. and Trembly, J. and Drabold, D. A.},
title = {Self-Assembly and the Properties of Micro-Mesoporous Carbon},
journal = {Journal of Chemical Theory and Computation},
volume = {20},
number = {4},
pages = {1753-1762},
year = {2024},
doi = {10.1021/acs.jctc.3c00394}
}

@article{ranganathan2015,
  title={Common pitfalls in statistical analysis: Odds versus risk},
  author={Ranganathan, Priya and Aggarwal, Rakesh and Pramesh, CS},
  journal={Perspectives in clinical research},
  volume={6},
  number={4},
  pages={222--224},
  year={2015},
  publisher={Medknow},
 doi={10.4103/2229-3485.167092 }
}

@Misc{OVITO,
  title        = {{Identify diamond structure}},
  howpublished = {OVITO},
  url          ={https://www.ovito.org/manual/reference/pipelines/modifiers/identify_diamond.html}
}

@article{GELUpaper,
      title={Gaussian Error Linear Units (GELUs)}, 
      author={Dan Hendrycks and Kevin Gimpel},
      year={2023},
      journal={arXiv},
      doi={10.48550/arXiv.1606.08415},
   }

@misc{GINEConvpaper,
      title={Strategies for Pre-training Graph Neural Networks}, 
      author={Weihua Hu and Bowen Liu and Joseph Gomes and Marinka Zitnik and Percy Liang and Vijay Pande and Jure Leskovec},
      year={2020},
      doi={10.48550/arXiv.1905.12265} 
}

@Misc{GINEConvpytorch,
  title        = {GINEConv},
  howpublished = {PyTorch Geometric},
  url          = {https://pytorch-geometric.readthedocs.io/en/stable/generated/torch_geometric.nn.conv.GINEConv.html},
}

@misc{DOLFINx,
  title     = {{DOLFINx}: the next generation {FEniCS} problem solving environment},
  author    = {Baratta, Igor A. and Dean, Joseph P. and Dokken, J{\o}rgen S. and Habera, Michal and Hale, Jack S. and Richardson, Chris N. and Rognes, Marie E. and Scroggs, Matthew W. and Sime, Nathan and Wells, Garth N.},
  doi       = {10.5281/zenodo.10447666},
  year      = {2023}
}

@Article{         PETSc,
  title         = {Parallel distributed computing using Python},
  author        = {Lisandro D. Dalcin and Rodrigo R. Paz and Pablo A. Kler and Alejandro Cosimo},
  journal       = {Advances in Water Resources},
  volume        = {34},
  number        = {9},
  pages         = {1124 - 1139},
  note          = {New Computational Methods and Software Tools},
  issn          = {0309-1708},
  doi           = {10.1016/j.advwatres.2011.04.013},
  year          = {2011}
}

@incollection{KrylovSolverBook,
title = {Chapter 4 The Krylov aspects},
author = {Charles George Broyden and Maria Teresa Vespucci},
series = {Studies in Computational Mathematics},
publisher = {Elsevier},
volume = {11},
pages = {77-103},
year = {2004},
booktitle = {Krylov Solvers for Linear Algebraic Systems},
issn = {1570-579X},
doi = {10.1016/S1570-579X(04)80005-3},
}

@article{UFL,
  title     = {Unified Form Language: A domain-specific language for weak formulations of partial differential equations},
  author    = {Alnaes, Martin S. and Logg, Anders and {\O}lgaard, Kristian B. and Rognes, Marie E. and Wells, Garth N.},
  journal   = {{ACM} Transactions on Mathematical Software},
  year      = {2014},
  volume    = {40},
  doi       = {10.1145/2566630},
}

@article{Deringer2021,
  title={Origins of structural and electronic transitions in disordered silicon},
  author={Deringer, Volker L and Bernstein, Noam and Cs{\'a}nyi, G{\'a}bor and Ben Mahmoud, Chiheb and Ceriotti, Michele and Wilson, Mark and Drabold, David A and Elliott, Stephen R},
  journal={Nature},
  volume={589},
  number={7840},
  pages={59--64},
  year={2021},
  doi={10.1038/s41586-020-03072-z},
  publisher={Nature Publishing Group UK London}
}

@article{Morrow2022,
    author = {Morrow, Joe D. and Deringer, Volker L.},
    title = {Indirect learning and physically guided validation of interatomic potential models},
    journal = {The Journal of Chemical Physics},
    volume = {157},
    number = {10},
    pages = {104105},
    year = {2022},
    month = {09},
    doi = {10.1063/5.0099929},
}

@article{Lubbers2018hipNN,
    author = {Lubbers, Nicholas and Smith, Justin S. and Barros, Kipton},
    title = {Hierarchical modeling of molecular energies using a deep neural network},
    journal = {The Journal of Chemical Physics},
    volume = {148},
    number = {24},
    pages = {241715},
    year = {2018},
    month = {03},
    issn = {0021-9606},
    doi = {10.1063/1.5011181}
}

@article{Batatia2025,
    author = {Batatia, Ilyes and Benner, Philipp and Chiang, Yuan and Elena, Alin M. and Kovács, Dávid P. and Riebesell, Janosh and Advincula, Xavier R. and Asta, Mark and Avaylon, Matthew and Baldwin, William J. and Berger, Fabian and Bernstein, Noam and Bhowmik, Arghya and Bigi, Filippo and Blau, Samuel M. and Cărare, Vlad and Ceriotti, Michele and Chong, Sanggyu and Darby, James P. and De, Sandip and Della Pia, Flaviano and Deringer, Volker L. and Elijošius, Rokas and El-Machachi, Zakariya and Fako, Edvin and Falcioni, Fabio and Ferrari, Andrea C. and Gardner, John L. A. and Gawkowski, Mikołaj J. and Genreith-Schriever, Annalena and George, Janine and Goodall, Rhys E. A. and Grandel, Jonas and Grey, Clare P. and Grigorev, Petr and Han, Shuang and Handley, Will and Heenen, Hendrik H. and Hermansson, Kersti and Ho, Cheuk Hin and Hofmann, Stephan and Holm, Christian and Jaafar, Jad and Jakob, Konstantin S. and Jung, Hyunwook and Kapil, Venkat and Kaplan, Aaron D. and Karimitari, Nima and Kermode, James R. and Kourtis, Panagiotis and Kroupa, Namu and Kullgren, Jolla and Kuner, Matthew C. and Kuryla, Domantas and Liepuoniute, Guoda and Lin, Chen and Margraf, Johannes T. and Magdău, Ioan-Bogdan and Michaelides, Angelos and Moore, J. Harry and Naik, Aakash A. and Niblett, Samuel P. and Norwood, Sam Walton and O’Neill, Niamh and Ortner, Christoph and Persson, Kristin A. and Reuter, Karsten and Rosen, Andrew S. and Rosset, Louise A. M. and Schaaf, Lars L. and Schran, Christoph and Shi, Benjamin X. and Sivonxay, Eric and Stenczel, Tamás K. and Sutton, Christopher and Svahn, Viktor and Swinburne, Thomas D. and Tilly, Jules and van der Oord, Cas and Vargas, Santiago and Varga-Umbrich, Eszter and Vegge, Tejs and Vondrák, Martin and Wang, Yangshuai and Witt, William C. and Wolf, Thomas and Zills, Fabian and Csányi, Gábor},
    title = {A foundation model for atomistic materials chemistry},
    journal = {The Journal of Chemical Physics},
    volume = {163},
    number = {18},
    pages = {184110},
    year = {2025},
    month = {11},
    issn = {0021-9606},
    doi = {10.1063/5.0297006}
}

@inbook{Lewis2004,
publisher = {John Wiley \& Sons, Ltd},
isbn = {9780470014165},
title = {Steady State Heat Conduction in One Dimension},
booktitle = {Fundamentals of the Finite Element Method for Heat and Fluid Flow},
chapter = {3},
pages = {38-101},
doi = {10.1002/0470014164.ch4},
year = {2004},
author ={Roland W. Lewis and Perumal Nithiarasu and Kankanhalli N. Seetharamu}
}

@Article{Hu2021MultiscaleReview,
author ="Hu, Ming and Yang, Zhonghua",
title  ="Perspective on multi-scale simulation of thermal transport in solids and interfaces",
journal  ="Phys. Chem. Chem. Phys.",
year  ="2021",
volume  ="23",
issue  ="9",
pages  ="5680-5680",
publisher  ="The Royal Society of Chemistry",
doi  ="10.1039/D1CP90037D"
}

@article{Kaminski2021Homo,
title = {Homogenization of heat transfer in fibrous composite with stochastic interface defects},
journal = {Composite Structures},
volume = {261},
pages = {113555},
year = {2021},
issn = {0263-8223},
doi = {10.1016/j.compstruct.2021.113555},
author = {Marcin Kamiński and Piotr Ostrowski},
}

@Article{Schutzeichel2021Homo,
AUTHOR = {Schutzeichel, Maximilian Otto Heinrich and Kletschkowski, Thomas and Monner, Hans Peter},
TITLE = {Microscale Thermal Modelling of Multifunctional Composite Materials Made from Polymer Electrolyte Coated Carbon Fibres Including Homogenization and Model Reduction Strategies},
JOURNAL = {Applied Mechanics},
VOLUME = {2},
YEAR = {2021},
NUMBER = {4},
PAGES = {739--765},
ISSN = {2673-3161},
DOI = {10.3390/applmech2040043}
}

@article{Peng2024ML,
  author  = {Peng, Xiang-Long and Fathidoost, Mozhdeh and Lin, Binbin and Yang, Yangyiwei and Xu, Bai-XiangXu},
  title   = {What Can Machine Learning Help with Microstructure-Informed Materials Modeling and Design?},
  journal = {MRS Bulletin},
  year    = {2024},
  volume  = {50},
  pages   = {61--79},
  doi     = {10.1557/s43577-024-00797-4}
}

@article{Li2022ML,
author = {Li, Man and Dai, Lingyun and Hu, Yongjie},
title = {Machine Learning for Harnessing Thermal Energy: From Materials Discovery to System Optimization},
journal = {ACS Energy Letters},
volume = {7},
number = {10},
pages = {3204-3226},
year = {2022},
doi = {10.1021/acsenergylett.2c01836}
}

@article{Keshav2025ML,
author = {Keshav, Sanath and Herb, Julius and Fritzen, Felix},
title = {Spectral Normalization and Voigt–Reuss net: A universal approach to microstructure-property forecasting with physical guarantees},
journal = {GAMM-Mitteilungen},
volume = {48},
number = {3},
pages = {e70005},
doi = {10.1002/gamm.70005},
year = {2025}
}

@article{XU2021SPDnn,
title = {Learning constitutive relations using symmetric positive definite neural networks},
journal = {Journal of Computational Physics},
volume = {428},
pages = {110072},
year = {2021},
issn = {0021-9991},
doi = {10.1016/j.jcp.2020.110072},
author = {Kailai Xu and Daniel Z. Huang and Eric Darve}
}

@article{Bangerth2017,
author = {Bangerth, Wolfgang and Kim, Imbunm and Sheen, Dongwoo and Yim, Jaeryun},
title = {On Hanging Node Constraints for Nonconforming Finite Elements using the Douglas--Santos--Sheen--Ye Element as an Example},
journal = {SIAM Journal on Numerical Analysis},
volume = {55},
number = {4},
pages = {1719-1739},
year = {2017},
doi = {10.1137/16M1071432},
}

@article{Yang2025,
    title = {Steady-state and transient thermal stress analysis using a polygonal finite element method},
    journal = {Finite Elements in Analysis and Design},
    volume = {251},
    pages = {104413},
    year = {2025},
    issn = {0168-874X},
    doi = {10.1016/j.finel.2025.104413},
    author = {Yang Yang and Mingjiao Yan and Zongliang Zhang and Dengmiao Hao and Xuedong Chen and Weixiong Chen}
    }

@article{BenAmor2019,
doi = {10.1088/1361-6528/ab29a6},
year = {2019},
month = {jul},
publisher = {IOP Publishing},
volume = {30},
number = {37},
pages = {375704},
author = {Ben Amor, Aymen and Djomani, Doriane and Fakhfakh, Mariam and Dilhaire, Stefan and Vincent, Laetitia and Grauby, Stéphane},
title = {Si and Ge allotrope heterostructured nanowires: experimental evaluation of the thermal conductivity reduction},
journal = {Nanotechnology},
}

@article{Vincent2018,
doi = {10.1088/1361-6528/aaa738},
year = {2018},
month = {feb},
publisher = {IOP Publishing},
volume = {29},
number = {12},
pages = {125601},
author = {Vincent, L and Djomani, D and Fakfakh, M and Renard, C and Belier, B and Bouchier, D and Patriarche, G},
title = {Shear-driven phase transformation in silicon nanowires},
journal = {Nanotechnology},
}

@article{Doerk2010,
author = {Doerk, Gregory S. and Carraro, Carlo and Maboudian, Roya},
title = {Single Nanowire Thermal Conductivity Measurements by Raman Thermography},
journal = {ACS Nano},
volume = {4},
number = {8},
pages = {4908-4914},
year = {2010},
doi = {10.1021/nn1012429}}

@article{ATOMSK,
title = {Atomsk: A tool for manipulating and converting atomic data files},
journal = {Computer Physics Communications},
volume = {197},
pages = {212-219},
year = {2015},
issn = {0010-4655},
doi = {10.1016/j.cpc.2015.07.012},
author = {Pierre Hirel}
}

@article{Haldane1956,
author = {Haldane, By J. B. S.},
title = {THE ESTIMATION AND SIGNIFICANCE OF THE LOGARITHM OF A RATIO OF FREQUENCIES},
journal = {Annals of Human Genetics},
volume = {20},
number = {4},
pages = {309-311},
doi = {10.1111/j.1469-1809.1955.tb01285.x},
year = {1956}
}

@article{Anscombe1956,
 ISSN = {00063444, 14643510},
 URL = {http://www.jstor.org/stable/2332926},
 author = {F. J. Anscombe},
 journal = {Biometrika},
 number = {3/4},
 pages = {461--464},
 publisher = {[Oxford University Press, Biometrika Trust]},
 title = {On Estimating Binomial Response Relations},
 volume = {43},
 year = {1956}
}

@book{zienkiewicz2005finite,
  title={The finite element method: Its Basis and Fundamentals},
  author={Zienkiewicz, Olgierd C and Taylor, Robert L and J. Z. Zhu},
  edition={6},
 page={229-253},
  year={2005},
  publisher={Elsevier Butterworth-Heinemann}
}

@book{yvonnet2019book,
  title={Computational homogenization of heterogeneous materials with finite elements},
  author={Yvonnet, Julien},
  volume={258},
  year={2019},
  doi={10.1007/978-3-030-18383-7},
  publisher={Springer}
}

@article{Chen2015,
title = {Evaluation of thermal conductivity of asphalt concrete with heterogeneous microstructure},
journal = {Applied Thermal Engineering},
volume = {84},
pages = {368-374},
year = {2015},
issn = {1359-4311},
doi = {10.1016/j.applthermaleng.2015.03.070},
author = {Jiaqi Chen and Miao Zhang and Hao Wang and Liang Li}
}

@article{Babuska1982,
author = {Babuska, Ivo and Szabo, Barna},
title = {On the rates of convergence of the finite element method},
journal = {International Journal for Numerical Methods in Engineering},
volume = {18},
number = {3},
pages = {323-341},
doi = {10.1002/nme.1620180302},
year = {1982}
}

@article{Gitman2007,
title = {Representative volume: Existence and size determination},
journal = {Engineering Fracture Mechanics},
volume = {74},
number = {16},
pages = {2518-2534},
year = {2007},
issn = {0013-7944},
doi = {10.1016/j.engfracmech.2006.12.021},
author = {I.M. Gitman and H. Askes and L.J. Sluys}
}

@article{Meng2025,
title = {Experimental and finite element analysis of thermal conductivity in Cu matrix composites reinforced with surface-modified Kovar particles},
journal = {Materials Chemistry and Physics},
volume = {333},
pages = {130392},
year = {2025},
issn = {0254-0584},
doi={10.1016/j.matchemphys.2025.130392},
author = {Tao Meng and Chaoqun Peng and Richu Wang and Zhiyong Cai}
}

@article{Basaula2022,
title = {Towards modeling thermoelectric properties of anisotropic polycrystalline materials},
journal = {Acta Materialia},
volume = {228},
pages = {117743},
year = {2022},
issn = {1359-6454},
doi = {10.1016/j.actamat.2022.117743},
author = {Dharma Basaula and Mohamad Daeipour and Lukasz Kuna and John Mangeri and Boris Feygelson and Serge Nakhmanson},
}

@article{Siddiqui2013,
title = {Finite element analysis of thermal conductivity and thermal resistance behaviour of woven fabric},
journal = {Computational Materials Science},
volume = {75},
pages = {45-51},
year = {2013},
issn = {0927-0256},
doi = {10.1016/j.commatsci.2013.04.003},
author = {Muhammad Owais Raza Siddiqui and Danmei Sun}
}

@article{twinGB,
author = {Isotta, Eleonora and Jiang, Shizhou and Bueno-Villoro, Ruben and Nagahiro, Ryohei and Maeda, Kosuke and Mattlat, Dominique Alexander and Odufisan, Alesanmi R. and Zevalkink, Alexandra and Shiomi, Junichiro and Zhang, Siyuan and Scheu, Christina and Snyder, G. Jeffrey and Balogun, Oluwaseyi},
title = {Heat Transport at Silicon Grain Boundaries},
journal = {Advanced Functional Materials},
volume = {34},
number = {40},
pages = {2405413},
doi = {10.1002/adfm.202405413},
year = {2024}
}

\end{document}


\maketitle

\begin{abstract} \bfseries \boldmath
\noindent Thermal transport in complex solids is governed by local structure, defects, and anisotropy, yet most continuum models still rely on oversimplified, homogenized conductivities. Here, we bridge atomistic and continuum descriptions by building finite element (FE) models directly from the site-projected thermal conductivity (SPTC), an atomic-level decomposition of the Green–Kubo thermal conductivity. We introduce a new toolkit, the ``Simulator Collection for Atomic-to-Continuum Scales (\texttt{SCACS})", which uses a graph neural network to predict SPTC on large atomic structures, coarse-grain these fields into anisotropic conductivity tensors, and embeds them into the heat-flow FE equation with a customized, anisotropy-aware adaptive mesh refinement scheme. Applied to silicon nanostructures, the resulting FE models act as representative volume elements, reproduce bulk conductivities, and capture interfacial and defect-driven anisotropy while maintaining thermodynamic consistency. Additionally, \texttt{SCACS} predicts experimental conductance trends and fields. This innovation demonstrates a novel and general route for transferring atomistic transport information into device-scale thermal simulations with physics-based approximations.
\end{abstract}

\section{\texttt{sptc-ai} Graph Implementation}\label{sec:features}

\subsection{Local Node Features}

We consider a dataset of atomistic structures represented as graphs \(\mathcal{G} = \{G^{(g)} = (V^{(g)}, E^{(g)})\}_{g=1}^G\), where each graph has node set \(V^{(g)}\) (atoms) and directed edge set \(E^{(g)}\) (neighbor pairs from a cutoff construction). For each node \(i \in V^{(g)}\), we construct a concatenated feature vector \(x_i \in \mathbb{R}^{F}\) assembled in a fixed order. The active feature blocks sum to a base dimensionality of \(F_{\mathrm{base}} = 27\). When SOAP (Smooth Overlap of Atomic Positions) descriptors are included, an additional 64 dimensions are appended, and when atomic-energy descriptors are also used, 2 further dimensions are added, yielding \(F = 91\) (SOAP only) or \(F = 93\) (SOAP + energy). The atomic-number (\(Z\)) embedding is omitted here because we focus solely on silicon, but remains available for multi-element systems; declared but unused fields (such as \(Z\)) are simply skipped without padding. One-hot and embedding blocks are not standardized, whereas scalar features are standardized (z-scored) using the mean and standard deviation computed from the training split. Targets are standardized
in the same way during training, and inverse transformations are applied during evaluation. Table~\ref{tab:node-features} summarizes all node-feature blocks used in the training stack, including their dimensionality, domain, and standardization policy. The column \texttt{struct.struct\_onehot} encodes the local structural phase for each node (amorphous, paracrystalline, crystalline, or crystalline with one or two vacancies).

\begin{table}[!tpbh]
\scriptsize
\setlength{\tabcolsep}{22pt}
\renewcommand{\arraystretch}{1.0}
\centering
\caption{Per-node feature blocks. ``Std?'' indicates whether the feature is standardized.}
\begin{tabular}{lclll}
\hline
\textbf{Block} (\texttt{graph.nodes[].}) & \textbf{Dim} & \textbf{Domain} & \textbf{Std?} & \textbf{Notes}\\
\hline
\multicolumn{5}{l}{\textbf{Structural (struct)}}\\
\texttt{struct.struct\_onehot} & 5 & simplex (\(\sum=1\)) & No & One-hot structural phase encoding\\
\texttt{struct.struct\_conf} & 1 & \([0,1]\) & Yes & Confidence of phase assignment \\
\texttt{struct.is\_gb} & 1 & \(\{0,1\}\) & Yes & Grain-boundary membership flag \\
\texttt{struct.vacancy\_cluster\_size} & 1 & \(\mathbb{Z}_{\ge 0}\) & Yes & Size of local vacancy cluster \\
\texttt{struct.Z\_embedding} & 32 & \(\mathbb{R}^{32}\) & No & Learned element embedding (optional) \\[3pt]

\multicolumn{5}{l}{\textbf{OVITO structure class descriptors (ovito)}}\\
\texttt{ovito.ovito\_type\_onehot} & 7 & simplex (\(\sum=1\)) & No & OVITO structure classifier (Table~\ref{tab:ovito-map}) \\
\texttt{ovito.csi\_depth} & 1 & \(\{-1,0,1,2\}\) & Yes & Local crystalline shell depth \\
\texttt{ovito.crystal\_fraction\_local} & 1 & \([0,1]\) & Yes & Local crystallinity score \\[3pt]

\multicolumn{5}{l}{\textbf{Statistical descriptors (stats)}}\\
\texttt{stats.coord\_num} & 1 & \(\mathbb{R}_{\ge 0}\) & Yes & Coordination number (bond-based) \\
\texttt{stats.first\_shell\_rms} & 1 & \(\mathbb{R}_{\ge 0}\) & Yes & RMS deviation of first shell \\
\texttt{stats.ptm\_error} & 1 & \(\mathbb{R}\) & Yes & Pattern-matching error metric \\
\texttt{stats.misorientation} & 1 & \(\mathbb{R}\) & Yes & Local orientation mismatch \\
\texttt{stats.mean\_bond} & 1 & \(\mathbb{R}\) & Yes & Mean bond length (within cutoff) \\
\texttt{stats.std\_bond} & 1 & \(\mathbb{R}_{\ge 0}\) & Yes & Std.\ of bond lengths \\
\texttt{stats.short\_bond\_frac} & 1 & \([0,1]\) & Yes & Fraction of short bonds \\
\texttt{stats.long\_bond\_frac} & 1 & \([0,1]\) & Yes & Fraction of long bonds \\
\texttt{stats.q4} & 1 & \(\mathbb{R}\) & Yes & Local Steinhardt order parameter \(q_4\) \\
\texttt{stats.q6} & 1 & \(\mathbb{R}\) & Yes & Local Steinhardt order parameter \(q_6\) \\[3pt]

\multicolumn{5}{l}{\textbf{SOAP descriptor (optional)}}\\
\texttt{soap.SOAP\_PCA64} & 64 & \(\mathbb{R}^{64}\) & No & PCA-reduced SOAP vector (train-fit basis) \\

\multicolumn{5}{l}{\textbf{Energy (optional)}}\\
\texttt{energy.E\_atom} & 1 & \(\mathbb{R}\) & Yes & Per-atom potential energy (if present) \\
\texttt{energy.has\_E\_atom} & 1 & \(\{0,1\}\) & No & Availability flag for \(E_\mathrm{atom}\) \\[3pt]
\hline
\end{tabular}
\label{tab:node-features}
\end{table}

For the \emph{OVITO} block in Table~\ref{tab:node-features}, If \(c_i \in \{0,\dots,6\}\) denote the structural class label
(obtained as the index of the one-hot vector) corresponding to the types described in the main text, and 
\(d_i \in \{-1,0,1,2\}\) is the coordination-shell index (CSI) depth, then the \texttt{csi\_depth} descriptor can be considered as a shell index that counts how many coordination shells are required to recognize a diamond-like environment (cubic or hexagonal). As summarized in Table~\ref{tab:ovito-map}, the \texttt{csi\_depth} provides a graded notion of crystalline context (how far one must look to see a diamond motif), while \texttt{ovito\_type\_onehot} encodes both the diamond family (cubic vs hexagonal) and the shell category (core, first shell, second shell).

\begin{table}[!tpbh]
\centering
\small
\caption{OVITO type order and CSI depth constraints.}
\begin{tabular}{clc}
\hline
\textbf{Class Index}& \textbf{Class Description}& \textbf{\texttt{csi\_depth}}\\
\hline
0 & Other & \(-1\) \\
1 & Cubic diamond & \(0\) \\
2 & Cubic diamond (1st neighbor) & \(1\) \\
3 & Cubic diamond (2nd neighbor) & \(2\) \\
4 & Hexagonal diamond & \(0\) \\
5 & Hexagonal diamond (1st neighbor) & \(1\) \\
6 & Hexagonal diamond (2nd neighbor) & \(2\) \\
\hline
\end{tabular}
\label{tab:ovito-map}
\end{table}

\subsection{Global Node Features}
The Feature-wise Linear Modulation (FiLM) adapter in \texttt{sptc-ai} uses a graph-level vector $\mathbf{g}_g \in \mathbb{R}^{d_{\mathrm{glob}}}$ constructed from a subset of the \texttt{graph.globals} fields in the dataset manifest. The vector is assembled in the order below (missing fields are filled with zeros), giving dimension $d_{\mathrm{glob}}=21$:
\begin{align*}
\mathbf{g}_g = \Big[
&\underbrace{q4_{\mathrm{mean}},\,q4_{\mathrm{std}},\,q6_{\mathrm{mean}},\,q6_{\mathrm{std}}}_{\text{bond-orientational order (4)}};\;
\underbrace{\mathrm{crystal\_frac\_mean}}_{\text{global crystal fraction (1)}};\;
\underbrace{h^{\mathrm{CSI}}_0,\,h^{\mathrm{CSI}}_1,\,h^{\mathrm{CSI}}_2,\,h^{\mathrm{CSI}}_3}_{\text{CSI depth histogram (4)}}; \underbrace{\mathrm{cn\_mean},\,\mathrm{cn\_std}}_{\text{coordination number stats (2)}};\;\nonumber\\
&\underbrace{\mathrm{vac\_size\_mean},\,\mathrm{vac\_size\_std}}_{\text{vacancy cluster stats (2)}};\;
\underbrace{L_x,\,L_y,\,L_z}_{\text{box lengths (3)}};\;
\underbrace{\mathbf{r}_{\mathrm{onehot}}^{\top}}_{\text{regime one-hot (4)}};\;
\underbrace{r_{\mathrm{conf}}}_{\text{regime confidence (1)}}
\Big]^{\!\top}\! \label{eq:film_globals_vec}
\end{align*}
Here $h^{\mathrm{CSI}}_k$ are the four-bin fractional count from CSI histogram for the model (bins for depths $-1,0,1,2$), 
$\mathbf{r}_{\mathrm{onehot}}\in\{0,1\}^4$ encodes the global structural regime (crystal, paracrystal, polycrystal, amorphous;
see Section~\ref{sec:regimeclass}). This design is intentionally similar to the local \texttt{struct.struct\_onehot} node descriptor and provides an additional consistency check for the implementation. The feature $r_{\mathrm{conf}}\in[0,1]$ is the confidence score of the regime classification. All other entries are scalar summaries from the the local node features: Steinhart bond-orientational ($q_4,q_6$) moments, global crystal fraction, coordination-number statistics, vacancy-cluster statistics, and box lengths $(L_x,L_y,L_z)$.

\subsection{Regime Classification}\label{sec:regimeclass}

Each structure is assigned to one of four structural regimes—crystal, polycrystalline, paracrystalline, or amorphous—based on its global crystalline fraction \((f_\mathrm{crys})\), grain count \((N_\mathrm{grain})\), and grain‐boundary fraction \((f_\mathrm{gb})\). Structures with a high crystalline fraction \((f_\mathrm{crys}\ge0.9)\), a single grain \((N_\mathrm{grain}\le1)\), and minimal grain‐boundary content \((f_\mathrm{gb}\le0.05)\) are labeled as crystal, with confidence increasing as \(f_\mathrm{crys}\) approaches 1 and \(f_\mathrm{gb}\) approaches 0. Systems that remain strongly crystalline \((f_\mathrm{crys}\ge0.6)\) but contain multiple grains \((N_\mathrm{grain}\ge2)\) and moderate boundary fractions \((f_\mathrm{gb}\le0.4)\) are classified as polycrystalline, where confidence grows with both the number of grains and the crystalline fraction. Configurations that are largely crystalline \((f_\mathrm{crys}\ge0.6)\) yet do not meet the stricter grain or boundary criteria are labeled paracrystalline, reflecting partial disorder or transitional states between amorphous and polycrystalline structures. All remaining systems, typically characterized by low crystalline fraction \((f_\mathrm{crys}<0.6)\), are labeled amorphous, with confidence increasing as crystalline order diminishes. The regime confidence score \((0\le C\le1)\) is computed from these margins to provide a continuous measure of classification certainty.

\section{Gaussian Occupancy Masking and Void Trimming}\label{sec:masking}

To remove void regions and spurious low-conductivity cells from the voxelized domain, the \texttt{sptc2fem} module constructs an auxiliary \emph{occupancy field} on the same fractional grid used for SPTC voxelization. We retain the notation
\(
N_{\mathrm{div}} = (n_x,n_y,n_z),
\)
fractional atom coordinates
\(
\boldsymbol{\xi}_a = M^{-1}\mathbf r_a \in [0,1)^3
\)
and voxel centers $\Xi^{\mathrm{cell}}_{ijk}$ and vertices $\Xi^{\mathrm{vert}}_{IJK}$ as defined in Section~\textcolor{blue}{4.4.2} in main text.

\subsection{Cell-based Gaussian occupancy}
We first define a Gaussian weight from atom $a$ to voxel $(i,j,k)$ using the same kernel as in the SPTC voxelization:
\begin{equation}
  \mathcal{G}^{\mathrm{cell}}_{ijk;a}
  = \exp\!\Big[
    -\frac{\big\|\Xi^{\mathrm{cell}}_{ijk}-\boldsymbol{\xi}_a\big\|_2^2}{2h_{\mathrm{frac}}^2}
  \Big]
\end{equation}
with $h_{\mathrm{frac}}$ given by the fractional-width definition already used for SPTC interpolation:
\begin{equation}
  h_{\mathrm{frac}} 
  = s \times \min\!\left(\tfrac{1}{n_x},\tfrac{1}{n_y},\tfrac{1}{n_z}\right),
  \qquad s = 0.75
\end{equation}
cf.\ the Gaussian smoothing in the voxelization equations.
Replacing the SPTC weights $\zeta^\beta_a$ in Equation \textcolor{blue}{47} by unity, we obtain a dimensionless cell occupancy: 
\begin{equation}
  \rho^{\mathrm{cell}}_{ijk}
  = \sum_{a=1}^{N_{\mathrm{at}}} \mathcal{G}^{\mathrm{cell}}_{ijk;a} \;;
  \qquad
  0 \le i < n_x,\ 0 \le j < n_y,\ 0 \le k < n_z
\end{equation}
Optionally, in a lattice-aware variant we replace the fractional squared distance in $\mathcal{G}^{\mathrm{cell}}_{ijk;a}$ by
\(
(\Xi^{\mathrm{cell}}_{ijk}-\boldsymbol{\xi}_a)^\mathsf{T} G\,(\Xi^{\mathrm{cell}}_{ijk}-\boldsymbol{\xi}_a)
/(2\sigma^2),
\)
with $G = M^\mathsf{T}M$ the Gram matrix and $\sigma$ a physical length scale constructed from the lattice vectors, but the logical structure of the mask remains unchanged.

\subsection{Vertex-based occupancy and induced cell mask.}
For some operations (e.g.\ isosurface extraction), it is convenient to define a vertex-based occupancy field:
\begin{equation}
  \rho^{\mathrm{vert}}_{IJK}
  = \sum_{a=1}^{N_{\mathrm{at}}}
    \exp\!\Big[
      -\frac{\big\|\Xi^{\mathrm{vert}}_{IJK}-\boldsymbol{\xi}_a\big\|_2^2}{2h_{\mathrm{frac}}^2}
    \Big]\,;
  \qquad
  0 \le I \le n_x,\ 0 \le J \le n_y,\ 0 \le K \le n_z
\end{equation}
A corresponding cell occupancy can then be induced by logical OR ($\bigvee$) over the eight corner vertices of each voxel:
\begin{equation}
\rho^{\mathrm{cell}}_{ijk} \ge \rho_{\mathrm{thr}}
\;\Longleftrightarrow\;
\bigvee_{I=i}^{i+1}
\bigvee_{J=j}^{j+1}
\bigvee_{K=k}^{k+1}
\left(\rho^{\mathrm{vert}}_{IJK} \ge \rho_{\mathrm{thr}}\right)
\end{equation}
where $\rho_{\mathrm{thr}}$ is a defined threshold. In practice we do not store this induced field explicitly; instead we construct the cell mask directly from the vertex field as described below.

\subsection{Occupancy thresholding}
Given either $\rho^{\mathrm{cell}}_{ijk}$ or $\rho^{\mathrm{vert}}_{IJK}$, we introduce an occupancy threshold $\tau_{\mathrm{occ}}\in[0,1]$ and define: 
\begin{align}
  \rho_{\max}^{\mathrm{cell}} &= \max_{i,j,k}\,\rho^{\mathrm{cell}}_{ijk}\\
  \rho_{\max}^{\mathrm{vert}} &= \max_{I,J,K}\,\rho^{\mathrm{vert}}_{IJK}
\end{align}
For the cell-based mode we define the occupancy mask as:
\begin{equation}
  m^{\mathrm{occ}}_{ijk}
  =
  \begin{cases}
    1, & \rho^{\mathrm{cell}}_{ijk} \ge \tau_{\mathrm{occ}}\,\rho_{\max}^{\mathrm{cell}}\\[4pt]
    0, & \text{otherwise}
  \end{cases}
\end{equation}
with the convention that if $\rho_{\max}^{\mathrm{cell}}=0$ we set $m^{\mathrm{occ}}_{ijk}=1$ for all cells, i.e.\ we keep the full brick (no trimming).

For the vertex-based mode we first define:
\begin{equation}
  m^{\mathrm{vert}}_{IJK}
  =
  \begin{cases}
    1, & \rho^{\mathrm{vert}}_{IJK} \ge \tau_{\mathrm{occ}}\,\rho_{\max}^{\mathrm{vert}}\\[4pt]
    0, & \text{otherwise}
  \end{cases}
\end{equation}
and then induce a cell mask by: 
\begin{equation}\label{eq:G_trim}
m^{\mathrm{occ}}_{ijk}
=
\bigvee_{I=i}^{i+1}
\bigvee_{J=j}^{j+1}
\bigvee_{K=k}^{k+1}
m^{\mathrm{vert}}_{IJK}.
\end{equation}
so a voxel is kept if at least one of its corner vertices is above the occupancy threshold.

\subsection{SPTC-based Conductivity Masking}
Independently of the occupancy, we also use the voxelized SPTC field to suppress cells with effectively vanishing conductivity. Using the nodal conductivities $k^\beta_{IJK}$ from Equation \textcolor{blue}{50} (either isotropic $\beta=\mathrm{d}$ or directional components), we construct a cell-wise scalar conductivity by averaging over the eight vertices of each voxel:
\begin{equation}\label{eq:K_trim}
  k^{\mathrm{cell}}_{ijk}
  = \frac{1}{8}
    \sum_{i=I-1}^{I}
    \sum_{j=J-1}^{J}
    \sum_{k=K-1}^{K}
    k_{IJK}^{\mathrm{d}}
\end{equation}
where $k^{\mathrm{d}}_{IJK}$ denotes the directionally averaged SPTC at vertex $(I,J,K)$. If we define
\(
k_{\max} = \max_{\substack{i,j,k}} k^{\mathrm{cell}}_{ijk},
\)
and choose a small relative cutoff $\varepsilon_k \ll 1$ (e.g.\ $\varepsilon_k = 10^{-12}$). We obtain the conductivity mask:
\begin{equation}
  m^{k}_{ijk}
  =
  \begin{cases}
    1, & k^{\mathrm{cell}}_{ijk} \ge \varepsilon_k\,k_{\max}\\[4pt]
    0, & \text{otherwise}
  \end{cases}
\end{equation}
with $m^k_{ijk}=0$ for all cells if $k_{\max}=0$.

\subsection{Final mask and mesh trimming}
The final binary mask used to trim the hexahedral mesh is the logical conjunction ($\wedge$) of Equation \ref{eq:G_trim} and \ref{eq:K_trim}: 
\begin{equation}
  m^{\mathrm{final}}_{ijk}
  = m^{\mathrm{occ}}_{ijk} \wedge m^{k}_{ijk}
\end{equation}
so a voxel is retained only if it is both sufficiently occupied (in the Gaussian sense) and has non-negligible SPTC-derived conductivity.

\begin{figure}[!h]
    \centering
    \includegraphics[width=\linewidth]{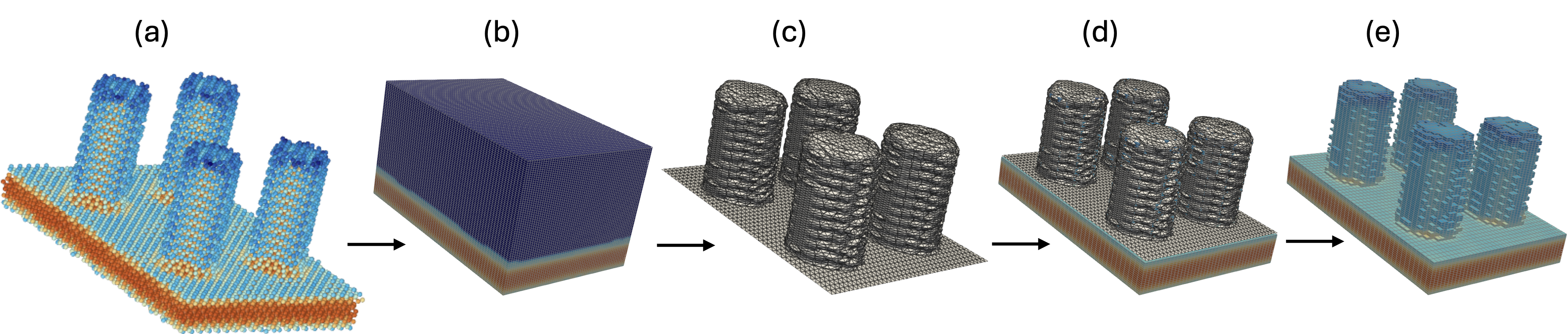}
    \caption{\textbf{Gaussian-form occupancy masking and void trimming in the Si nanopillar structure.} 
    (a) The atomic-scale model. (b) The structure without applying any mask for void trimming (c) isosurface resulting from masking and void trimming (d) isosurface overlay on the Si nanopillar structure and (e) the final continuum-scale structure for finite element analysis.  }
    \label{fig:SFig_sptc2fem_masking}
\end{figure}

On the structured brick, each index triplet $(i,j,k)$ corresponds to one hexahedral element in the FE mesh. The \texttt{sptc2fem} preprocessing step discards all elements with $m^{\mathrm{final}}_{ijk}=0$, collects the subset of vertices referenced by the remaining elements, renumbers them to obtain a contiguous node set, and restricts all nodal fields (e.g.\ $k^\beta_{IJK}$) to this trimmed mesh. The resulting domain excludes void regions and numerically problematic low-conductivity cells, while preserving the SPTC-consistent anisotropy encoded in the voxelized conductivity field $\mathbf{K}$ used in the FE formulation. We show an illustration of the masking and void trimming method implemented for the silicon nanopillar structure  in Figure \ref{fig:SFig_sptc2fem_masking}.

\section{Dataset Description}
We provide a dataset of the \texttt{sptc-ai} predicted SPTC models (extended \texttt{.xyz}) and the \texttt{sptc2fem} coarse-grain mesh (\texttt{.xdmf} visualization files with associated \texttt{.h5} files) discussed in the main manuscript \cite{Zenodo}.  In addition to the 8 models presented, we also include a single-pillar nanopillar structure. The name and description of the files are provided in Table \ref{tab:aux_mat_description}. The atomic structures are best visualized using OVITO \cite{OVITO} and the FE mesh is best visualized using ParaView \cite{ParaView}.

\begin{table}[!tpbh]
\scriptsize
\setlength{\tabcolsep}{22pt}
\renewcommand{\arraystretch}{1.0}
\centering
\caption{Description of the auxiliary supplemetary material}
\begin{tabular}{ll}
\hline
\textbf{Folder/Files} & \textbf{Description}\\
\hline
\textbf{sptcai/}              & \textbf{Atomic structure for:}\\
\texttt{1\_AC.pred.xyz}       & 2022-atom amorphous-crystalline Si structure (Figure \textcolor{blue}{2a}) \\
\texttt{2\_GB.pred.xyz}       & 223,930-atom Si twin-grain-boundary nanowire (Figure \textcolor{blue}{2c})  \\
\texttt{3\_Pxtal.pred.xyz}    & 134,923-atom polycrystalline Si nanowire (Figure \textcolor{blue}{2d}) \\
\texttt{4\_NP1.pred.xyz}      & 26,191-atom Si (4-pillar) nanopillar structure (Figure \textcolor{blue}{2e}) \\
\texttt{5\_NP2.pred.xyz}      & Additional 6531-atom Si (single-pillar) nanopillar structure (not in manuscript) \\
\texttt{6\_Ptype600.pred.xyz} & 3C/2H Si polytype structure @ 600$^\circ$C (Figure \textcolor{blue}{5a} LEFT) \\
\texttt{7\_Ptype700.pred.xyz} & 3C/2H Si polytype structure @ 700$^\circ$C (Figure \textcolor{blue}{5b} LEFT) \\
\texttt{8\_Ptype900.pred.xyz} & 3C/2H Si polytype structure @ 900$^\circ$C (Figure \textcolor{blue}{5c} LEFT) \\
\texttt{9\_Ptype100.pred.xyz} & 3C/2H Si polytype structure @ 1000$^\circ$C (Figure \textcolor{blue}{5d} LEFT) \\[3pt]
\textbf{sptc2fem/}              & \textbf{Finite element mesh for:}\\
\texttt{1\_AC\_mesh.xdmf(.h5)}       & 2022-atom amorphous-crystalline Si structure (Figure \textcolor{blue}{3a}) \\
\texttt{2\_GB\_mesh.xdmf(.h5)}       & 223,930-atom Si twin-grain-boundary nanowire (Figure \textcolor{blue}{4a})  \\
\texttt{3\_Pxtal\_mesh.xdmf(.h5)}    & 134,923-atom polycrystalline Si nanowire (Figure \textcolor{blue}{4b,c}) \\
\texttt{4\_NP1\_mesh.xdmf(.h5)}      & 26,191-atom Si (4-pillar) nanopillar structure (Figure \textcolor{blue}{4d}) \\
\texttt{5\_NP2\_mesh.xdmf(.h5)}      & Additional 6531-atom Si (single-pillar) nanopillar structure (not in manuscript) \\
\texttt{6\_Ptype600\_mesh.xdmf(.h5)} & 3C/2H Si polytype structure @ 600$^\circ$C (Figure \textcolor{blue}{5a} RIGHT) \\
\texttt{7\_Ptype700\_mesh.xdmf(.h5)} & 3C/2H Si polytype structure @ 700$^\circ$C (Figure \textcolor{blue}{5b} RIGHT) \\
\texttt{8\_Ptype900\_mesh.xdmf(.h5)} & 3C/2H Si polytype structure @ 900$^\circ$C (Figure \textcolor{blue}{5c} RIGHT) \\
\texttt{9\_Ptype100\_mesh.xdmf(.h5)} & 3C/2H Si polytype structure @ 1000$^\circ$C (Figure \textcolor{blue}{5d} RIGHT) \\[3pt]
\hline
\end{tabular}
\label{tab:aux_mat_description}
\end{table}

\section{Figures}

\begin{figure}[!h]
    \centering
    \includegraphics[width=\linewidth]{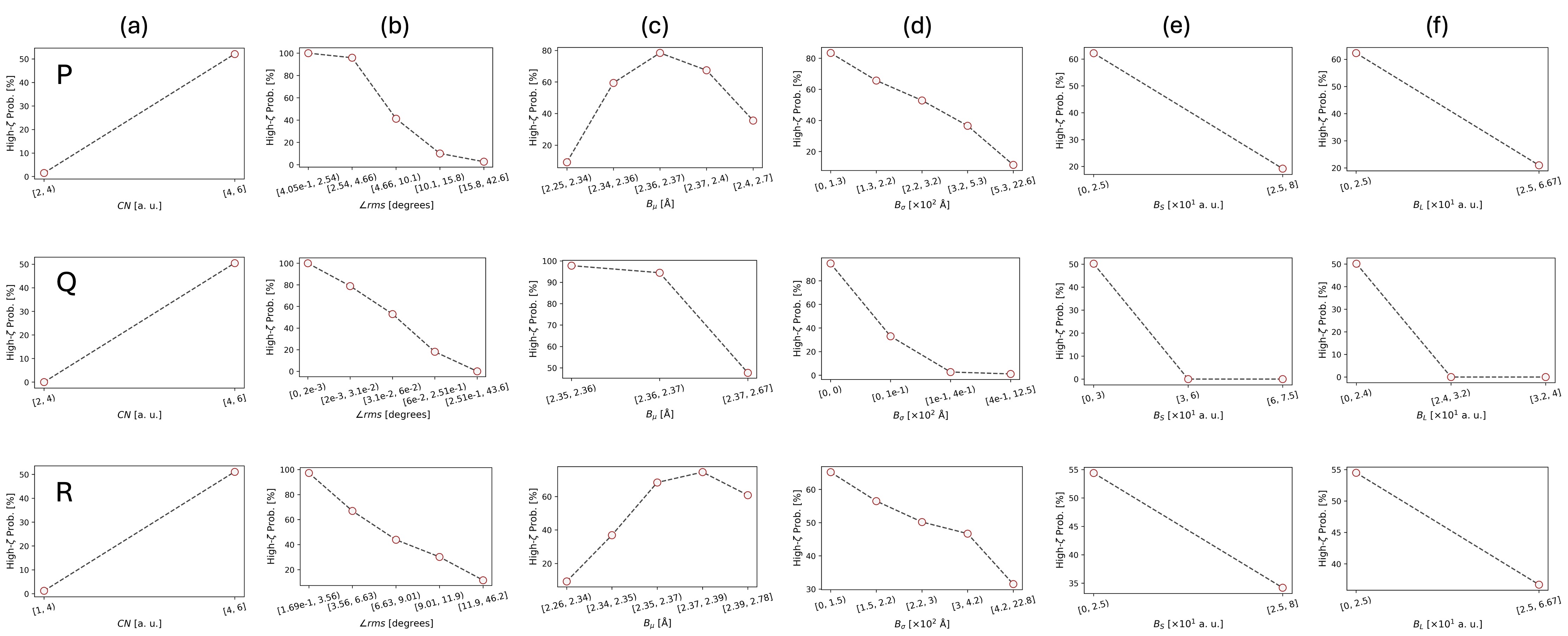}
    \caption{\textbf{Statistical Descriptors for \emph{P}, \emph{Q}, and \emph{R} Si structure groups used in training \texttt{sptc-ai}.} 
  Empirical probability of high-SPTC outcome from binned distribution of the continuous structural descriptors, including (a) the coordination number (CN), (b) the root-mean-square deviation from the ideal tetrahedral angle ($\angle\mathrm{rms}$), (c) the mean (B$_\mu$) and (d) standard deviation (B$_\sigma$) of bond lengths, and the (e) fractions of short (B$_\mathrm{S}$) and (f) long (B$_\mathrm{L}$) bonds. The top, middle and bottom rows are for the \emph{P}, \emph{Q} and \emph{R} structure groups, respectively.}
    \label{fig:SFig_sptcai_PQRdescriptors}
\end{figure}

\begin{figure}[!h]
    \centering
    \includegraphics[width=\linewidth]{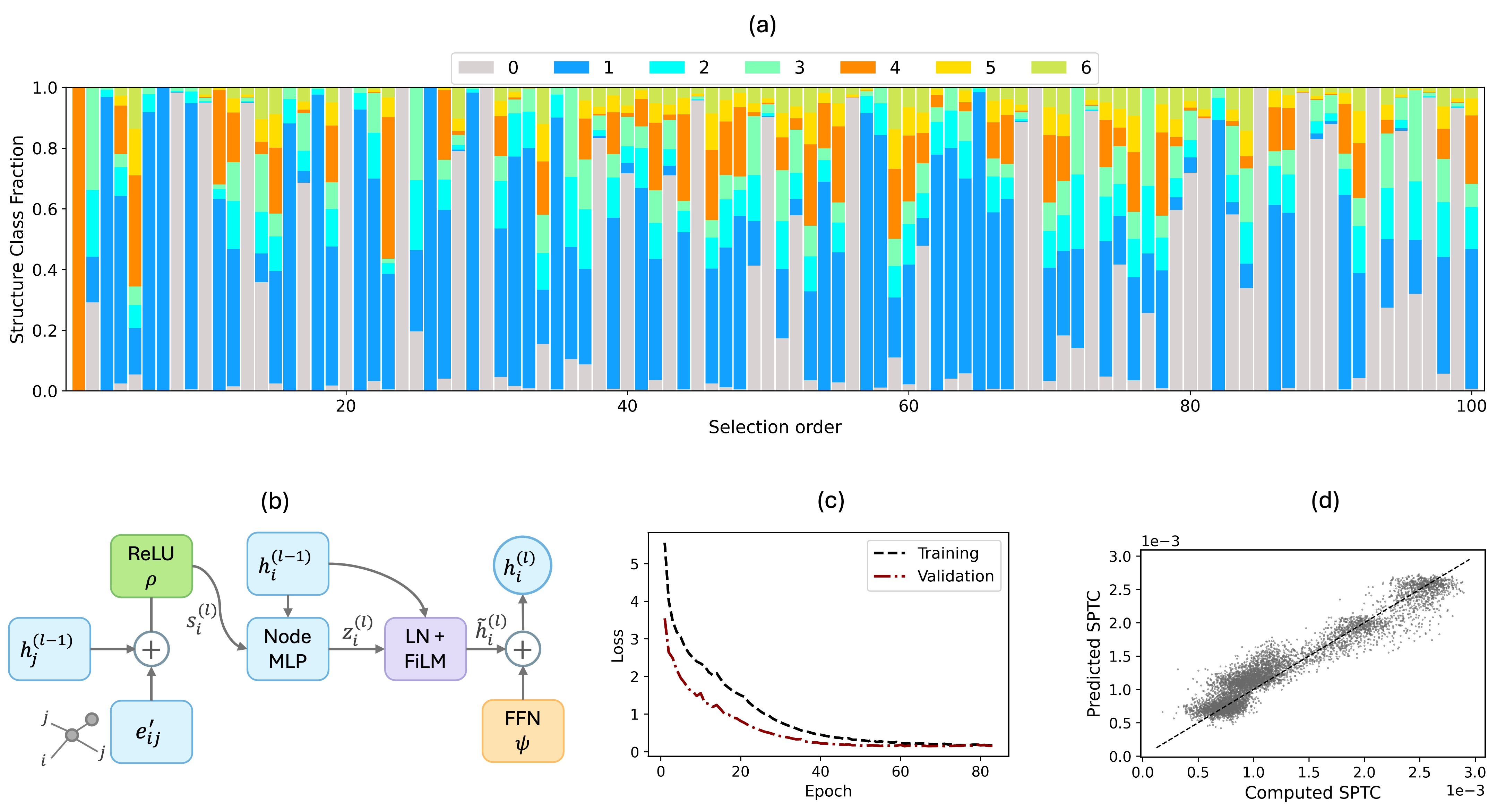}
    \caption{\textbf{Architecture and training of \texttt{sptc-ai}.} 
    (a) Distribution of local atomic environments for the top 100 unique Si structures. Colors denote structural labels—amorphous (0; gray), cubic diamond (1; blue) and its first (2; cyan) and second (3; green) neighbors, and hexagonal diamond (4; orange) with its first (5; yellow) and second (6; lime) neighbors. 
    (b) Schematic representation of the \texttt{sptc-ai} graph neural network architecture. MLP, LN, and FFN abbreviates multi-layer perceptron, layer normalization, and feed-forward network, respectively. (c) Loss convergence plot for training and validation. (d) Parity plot for predicted and computed (true) SPTC on an unseen test set.}
    \label{fig:SFig_sptcai_training}
\end{figure}

\clearpage 
\begin{figure}[!t]
    \centering
    \includegraphics[width=\linewidth]{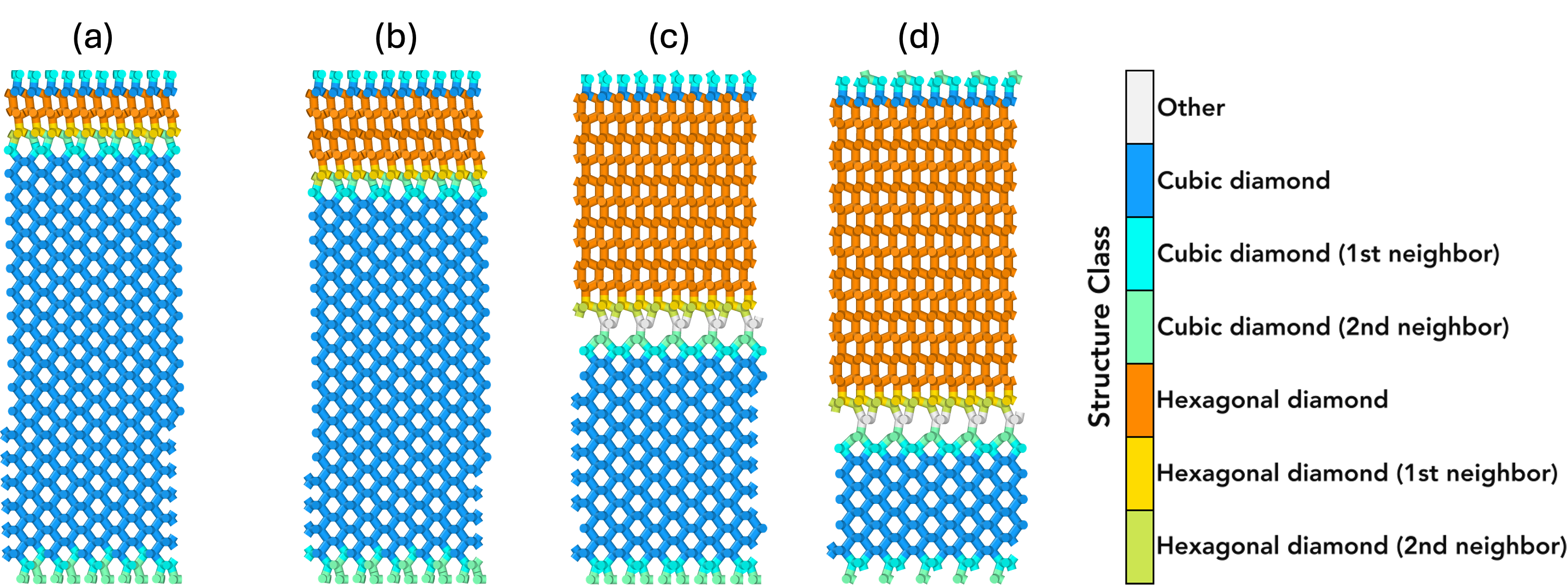}
    \caption{\textbf{Silicon polytype structure class.} 
    (a--d) The structure class representation for the 3C/2H nanowire models at annealing temperatures of 600, 700, 900, and 1000~$^\circ$C.  }
    \label{fig:SFig_sptc2fem_polytypeStructClass}
\end{figure}

\bibliography{references} 
\bibliographystyle{sciencemag}